\documentclass[a4paper]{article}
\pdfoutput=1

\usepackage{physics,bm,enumerate,multirow,amsmath,amssymb,hyperref,graphicx,float,authblk,tabularx,cite}
\usepackage[font=small,labelfont=bf]{caption}

\newcommand{\reffig}[2]{\hyperref[#1]{figure~\ref*{#1}#2}}
\newcommand{\refeq}[1]{\hyperref[#1]{equation~(\ref*{#1})}}
\newcommand{\refeqsystem}[1]{\hyperref[#1]{equations~(\ref*{#1})}}
\newcommand{\refeqshort}[1]{\hyperref[#1]{Eq~(\ref*{#1})}}
\newcommand{\refeqs}[2]{\hyperref[#1]{equations~(\ref*{#1})} and \hyperref[#2]{(\ref*{#2})}}

\newcommand{\refFig}[2]{\hyperref[#1]{Figure~\ref*{#1}#2}}
\newcommand{\refEq}[1]{\hyperref[#1]{Equation~(\ref*{#1})}}
\newcommand{\refEqs}[2]{\hyperref[#1]{Equations~(\ref*{#1})} and \hyperref[#2]{(\ref*{#2})}}

\newcommand{\refelement}[2]{\hyperref[#2]{#1~\ref*{#2}}}

\newcommand{\refsubsection}[2]{\hyperref[#2]{section~\ref*{#1}\ref*{#2}}}
\newcommand{\refSubsection}[2]{\hyperref[#2]{Section~\ref*{#1}\ref*{#2}}}

\newcommand{\intinf}{\int_{-\infty}^\infty}
\newcommand{\dz}{\:\text{d}z}
\newcommand{\etal}{{\em et al. }}
\newcommand{\bigok}[1]{\mathcal{O}(\kappa^{#1})}

\newenvironment{keywords}
    {\vfill\hrule\vspace{0.2cm}
    	\begin{tabular}{p{0.2\hsize}p{0.7\hsize}}
    	\textbf{Keywords:} &
    }
    {
    \end{tabular}
    \vspace{0.2cm}
    \clearpage
    }

\begin{document}

\title{Reversible signal transmission in an active mechanical metamaterial}

\author[1,2]{Alexander P Browning}
\author[3]{Francis G Woodhouse}
\author[1\footnote{Corresponding author. E-mail: matthew.simpson@qut.edu.au}]{Matthew J Simpson}
\affil[1]{Mathematical Sciences, Queensland University of Technology, Brisbane, Australia}
\affil[2]{ARC Centre of Excellence for Mathematical and Statistical Frontiers, QUT, Australia}
\affil[3]{Mathematical Institute, University of Oxford, Oxford, UK}

\maketitle

\begin{abstract}

\noindent 
Mechanical metamaterials are designed to enable unique functionalities, but are typically limited by an initial energy state and require an independent energy input to function repeatedly. Our study introduces a theoretical active mechanical metamaterial that incorporates a biological reaction mechanism to overcome this key limitation of passive metamaterials. Our material allows for reversible mechanical signal transmission, where energy is reintroduced by the biologically motivated reaction mechanism. By analysing a coarse grained continuous analogue of the discrete model, we find that signals can be propagated through the material by a travelling wave. Analysis of the continuum model provides the region of the parameter space that allows signal transmission, and reveals similarities with the well-known FitzHugh-Nagumo system. We also find explicit formulae that approximate the effect of the timescale of the reaction mechanism on the signal transmission speed, which is essential for controlling the material. 

\end{abstract}


\section{Introduction}
\label{s:Introduction}

Mechanical metamaterials are artificially constructed and have mechanical properties defined by their structure \cite{Bertoldi:2017bb}. Simple metamaterials consist of a one, two, or three-dimensional array of elements connected by links \cite{Bertoldi:2017bb,Paulose:2015hd,Hwang:2018ju} that may be elastic \cite{Nadkarni2016,Raney2016,Deng:2018jk,Kochmann:2017hz}, magnetic \cite{Dudek:2018ce,SerraGarcia:2018gl} or electrostatic \cite{Nadkarni2016}.   Mechanical metamaterials are highly tuneable \cite{Silverberg:2014uj,Paulose:2015dc,Turco:2017hc} and by altering the structure of these elements, and the properties of the links, materials have been developed that selectively transmit signals \cite{Fang:2017hi,Deng:2017gh}, behave as logic gates \cite{Raney2016,Ion:2017bo} or buckle after the application of external stimulus \cite{Paulose:2015hd}.   There are many recent studies that experimentally realise simple mechanical metamaterials \cite{Matlack:2016fp,Deng:2017gh,Deng:2018jk,Chen:2017uy,Chen:2017fl,Baardink2018}.   An advantage of these designs is that they are often well suited to utilise three-dimensional printing technology \cite{Matlack:2016fp,Raney2016,Chen:2017fl,Ding:2017ej,Hwang:2018ju}, however a common theme among existing metamaterials is that they generally require an external source of energy be provided in order to power their functions \cite{Wehner:2014be,Raney2016}. Many existing technologies can be thought of as \textit{static} or \textit{inactive} in the sense that they are limited by a fixed initial energy state, and are only able to respond to a finite number of stimuli before the manual introduction of external energy.

\begin{figure}[!t]
	\centering
	\includegraphics[width=0.9\textwidth]{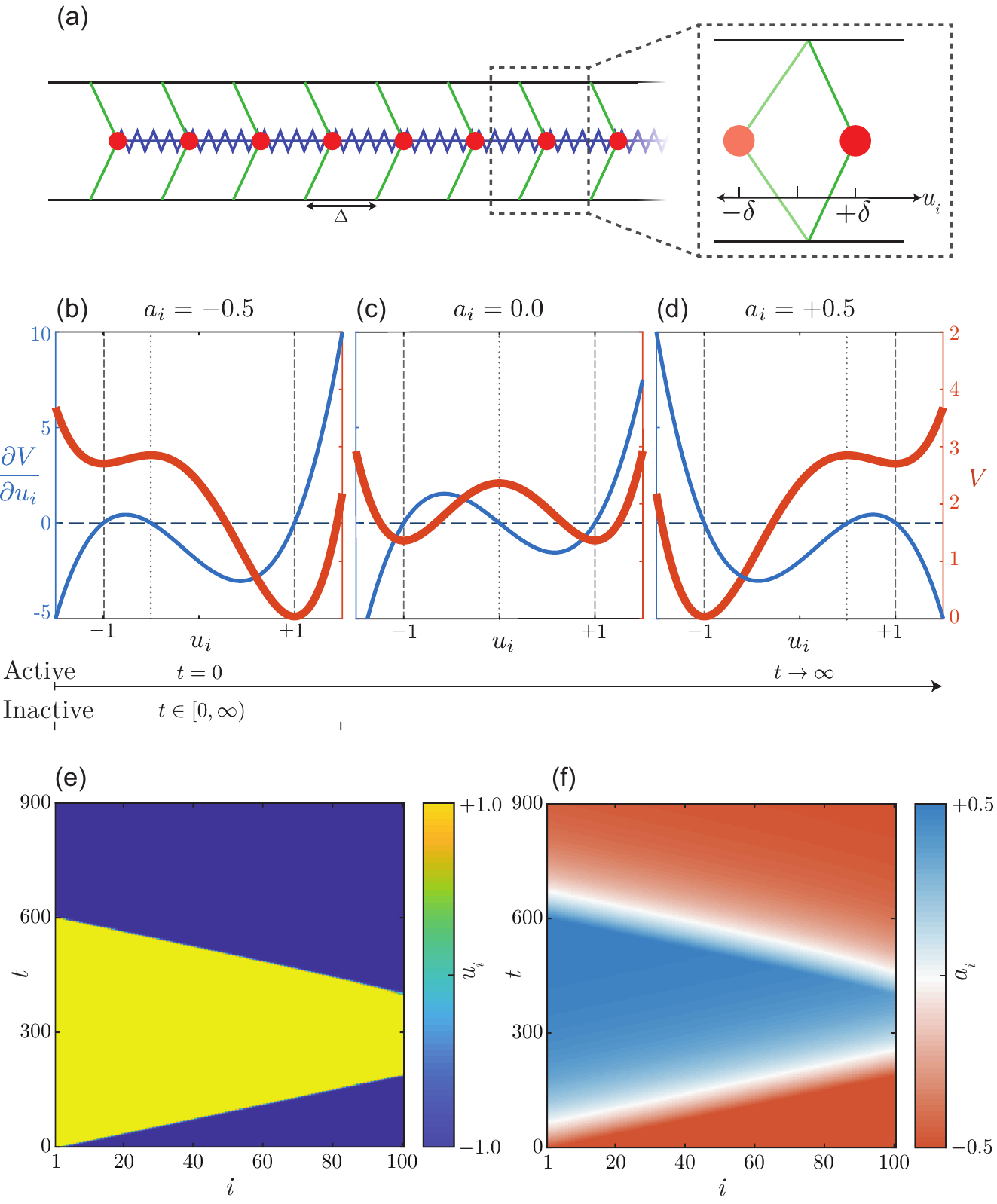}
	\caption{(a) Schematic of the one dimensional metamaterial \cite{Raney2016}. Each element (red), of mass $m$, has a natural spacing of $\Delta$. Each mass is connected to an external wall by a bistable elastic element (green) and to neighbouring elements by linear elastic springs (blue). The inset shows both stable states for each element, $u_i(t) = \pm \delta$ where $i$ is the mass index. (b) An example bistable potential function (red) and its' derivative (blue) for $a_i = -0.5$ and $\delta = 1$ in the inactive model, or the active model at $t = 0$. (b)-(d) The effect of the biological reaction mechanism, which resets the potential function. Note that (d) corresponds to a reflection of (b) about $u_i = 0$. For each plot the stable steady states (dashed grey) and unstable steady state (dotted grey) are shown. (e)-(f) Signal propagation through this material showing (e) the displacement of each element, $u_i(t)$; and, (f) the biological reaction, $a_i(t)$. The signal was initiated by moving the first element from a displacement of $-\delta$ to $\delta$ at $t = 0$, and was retransmitted by moving the last element from a displacement of $\delta$ to $-\delta$ at $t = 400$.  We note the original signal could also be initiated from the right boundary by moving the last element from a displacement of $-\delta$ to $\delta$, or indeed anywhere in the domain.   Parameters used are $m = 1$ g, $k = 1$ g m/s$^2$, $\gamma = 1$ g/s, $\Delta = 0.002$ m, $\delta = 1$ m, $\epsilon = 0.01$ /s, $\eta = 2$, $v = 1$ g/(m$^2$s$^2$) and $N = 101$ masses.}
	\label{Fig-1-Schematic}
\end{figure}

Recent mathematical and experimental work examines properties of a class of one-dimensional bistable metamaterials \cite{Nadkarni2016,Raney2016}. These systems comprise elements, each consisting of a mass connected to an external wall by a set of elastic elements that produce a bistable elastic potential. Individual elements are arranged in a one-dimensional lattice and interconnected by linear springs (\reffig{Fig-1-Schematic}{a}). These elements may be tuned so that the elastic potential energy function is asymmetric, resulting in both a high and low potential energy stable configuration for each element (\reffig{Fig-1-Schematic}{b}).   The system can therefore be designed so that an external stimulus, for example the change of a single node from the high to low potential energy stable state, can trigger a change in element configuration through the entire lattice \cite{Raney2016}.  This change is the transmission of a mechanical signal powered by stored elastic potential energy.

A limitation of this mechanical regime is that the system must receive an external energy input before the transmission of an additional signal \cite{Hwang:2018ju,Raney2016}.  Stiffness grading has been shown to overcome this limitation by exploiting a symmetric potential function \cite{Hwang:2018ju}, however these techniques may not allow the propagation of waves in systems with nonzero damping. The material in its current state is reset by manually moving each element back into its high potential energy configuration  \cite{Raney2016}. Our study introduces a theoretical biologically inspired mechanism that automatically resets each element to a high potential energy state, allowing the transmission of further signals. Many recent studies introduce the idea of manipulating biological subsystems in materials \cite{Wiktor:2011hj,Miniaci:2016cq,Jiang:2019fo}, or discuss behaviours that arise in \textit{active matter} systems, where biological systems exert mechanical forces \cite{Woodhouse:2018cw}.   Some systems are biologically inspired \cite{Wehner:2014be,Miniaci:2016cq,Jiang:2019fo} where properties of the metamaterial are designed to mimic a biological phenomenon, and some systems exploit the properties of biological subsystems to produce new behaviours in the material \cite{Wiktor:2011hj}.   One possibility for our mechanism is to exploit actin filaments in eukaryotic cells \cite{Pollard:1986cz,Blanchoin:2014jr,Kumar:2017hw,Nicolau:2016kf} to convert energy provided to the cells as nutrients through chemical hydrolysis \cite{Rouvala:WNdL2lqT} into mechanical energy which can reset the bistable elements to a high potential energy configuration.  The application of actin filaments in nanotechnology is well studied \cite{Kumar:2017hw} and their exploitation in metamaterials has been previously suggested \cite{Rouvala:WNdL2lqT,Nicolau:2016kf}.

The biological reaction mechanism we introduce is designed to react to changes in the displacement of individual elements and respond by inducing elastic potential energy back into each element.   Mathematically, the effect of this process is to reset the potential function so that each element eventually reverts to a high potential energy state, as shown in \reffig{Fig-1-Schematic}{b-d}.  We present a mathematical characterisation of this reaction mechanism and  explore how  the timescale of the reaction mechanism affects the ability of the system to transmit signals.  We find that signal transmission through a coarse-grained description of the material takes the form of a travelling wave. Using a travelling wave model, we find explicit formulae that bound the parameter space for which signal transmission can occur, and approximate the effect the timescale of the reaction mechanism has on the signal transmission speed. We lay the foundation for future work on this system where the metamaterial can be tuned to produce useful new behaviours. The results we provide quantify the trade-off between the signal transmission speed and the timescale of the biological response, which are essential for control and tuning of the material. For clarity, throughout this work we refer to the system without the reaction mechanism as the \textit{inactive} system, and the system that includes the biological reaction mechanism we introduce as the \textit{active} system.

In \refelement{section}{s:MathematicalModel} we present a mathematical model that describes the discrete active mechanical system. Following this, we take a continuous limit of the discrete model \cite{Nadkarni2016} with which we qualitatively explore the effect of the reaction mechanism on the ability of the system to transmit mechanical signals. We find evidence of travelling waves in the continuous model, where the wavespeed corresponds to the signal transmission speed. In  \refelement{section}{s:TravellingWaveModel}  we solve for the wavespeed and shape in the case where the reaction mechanism is excluded. This analysis is then extended to explore the effect that the reaction mechanism has on the wave speed and shape by taking a singular perturbation expansion (\refsubsection{s:TravellingWaveModel}{ss:PerturbationSolution}) and applying an energy conservation argument (\refsubsection{s:TravellingWaveModel}{ss:EnergyTransport}). Finally, in \refelement{section}{s:Discussion} we discuss and summarise our results, and outline future work involving our active metamaterial.

\section{Mathematical model}
\label{s:MathematicalModel}

The metamaterial presented by Raney \etal \cite{Raney2016} consists of $N$ bistable elements of mass $m$, interconnected by $N-1$ linear springs of stiffness $k$, and to two external walls by a pair of elastic elements, with a separation of $\Delta$. A schematic of this physical system is shown in \reffig{Fig-1-Schematic}{a}. Denoting the displacement of the $i$th mass relative to the mean of its two steady states as $u_i(t)$, the state of the discrete system is governed by
	\begin{equation}\label{eq:DiscreteODE}
	\begin{aligned}
		&0 = m\frac{\text{d}^2u_i}{\text{d}t^2} - k(u_{i+1} - u_i - \Delta) + \gamma \frac{\text{d}u_i}{\text{d}t} + \pdv{V}{u_i}, \: &&i = 1,\\
		&0 = m\frac{\text{d}^2u_i}{\text{d}t^2} - k(u_{i+1} - 2u_i + u_{i-1}) + \gamma \frac{\text{d}u_i}{\text{d}t} + \pdv{V}{u_i}, \: &&i = 2,...,N-1,\\
		&0 = m\frac{\text{d}^2u_i}{\text{d}t^2} - k(u_{i} - u_{i-1} + \Delta) + \gamma \frac{\text{d}u_i}{\text{d}t} + \pdv{V}{u_i}, \: &&i = N,
	\end{aligned}
	\end{equation}
where $\gamma$ is a damping parameter and $V(u_i,a_i)$ describes the potential energy of the bistable elements that connect each mass to the external wall \cite{Raney2016,Nadkarni2016}.

In this study, we choose $V(u_i,a_i)$ to be a quartic \cite{Nadkarni2016} defined in terms of its derivative,
	\begin{equation}\label{eq:DiscreteODE-V}
		\pdv{V}{u_i} = v(u_i^2 - \delta^2)(u_i - a_i),
	\end{equation}
where $v$ describes the stiffness of the bistable elements and relates to the size of the energy gap between high and low potential energy configurations, $u_i = \pm \delta$ are the stable fixed points of $V(u_i,a_i)$ and $u_i = a_i \in [-\delta,\delta]$ is the unstable fixed point which governs the symmetry of $V(u_i,a_i)$ (\reffig{Fig-1-Schematic}{b}). For this choice of potential energy function, $u_i = \text{sign}(a_i)\delta$ corresponds to an element in the high potential energy configuration and $u_i = -\text{sign}(a_i)\delta$ corresponds to an element in the low potential energy configuration (\reffig{Fig-1-Schematic}{b}).

In our study we allow the symmetry of the potential function to vary in reaction to changes in the displacement, $u_i(t)$, by allowing $a_i = a_i(t)$ and enforcing
	\begin{equation}
		\dv{a_i}{t} = \epsilon\left(\dfrac{u_i}{\eta} - a_i\right).
	\end{equation}
Physically, $a_i$ represents a biological subsystem that receives energy from external sources and induces it into the material at a rate proportional to $\epsilon$. The parameter $\eta > 1$ determines the extrema of the reaction parameter so that $a_i(t) \in [-\delta/\eta,\delta/\eta]$. This implementation means that if a user transmits a signal through the material by changing the displacement of a node (\reffig{Fig-1-Schematic}{e}), and waits a period of time of $\mathcal{O}(\epsilon^{-1})$ after the signal reaches the end of the domain, all elements of the system will have reverted to their high potential energy states. This resetting process of $a_i(t)$, and by extension $V(u_i,a_i)$, is shown for a single element \reffig{Fig-1-Schematic}{b-d}, and throughout the material in \reffig{Fig-1-Schematic}{f}. The discrete system is similar to other fast-slow bistable systems, such as the FitzHugh-Nagumo model \cite{FitzHugh1955,Rubin:2008kw,Beck:2008bx}.  In this context, we consider that the displacement function, $u_i(t)$, undergoes an excitable excursion in phase space in response to an external stimulus, and the variable representing the biological response, $a_i(t)$, behaves as a linear recovery variable.  In addition to the results in \reffig{Fig-1-Schematic}{e}, we reproduce results in the supporting material in the case the second signal is initiated too early, so propagation can't occur.

If the length of the material is large relative to the separation of each mass, $\Delta$, we can describe the material with a coarse grained continuous model \cite{Baker:2018ji}.  To derive a continuous description of the system described by \refeqsystem{eq:DiscreteODE}, we consider a material of fixed length, $L = (N-1)\Delta$, and take the limit $N \rightarrow \infty$ so that $\Delta \rightarrow 0$.  Following this, we define field functions $u(x,t)$ and $a(x,t)$ that describe $u_i(t)$ and $a_i(t)$, respectively, for $x = (i-1)\Delta \in (0,L)$. When taking a continuous limit of the discrete system we require that the macroscopic quantities in the discrete model remain $\mathcal{O}(1)$ for physical reasons\cite{Nadkarni2016}.  To do this, we replace unit quantities $m$, $v$ and $\gamma$ with density quantities, $\rho = m/\Delta$, $\hat{v} = v/\Delta$ and $\hat\gamma = \gamma / \Delta$, and scale the connecting spring force, $\hat{k} = \Delta k$.

Dividing \refeq{eq:DiscreteODE} by $\Delta$ and taking the limit $\Delta \rightarrow 0$ results in the continuous model,
	\begin{align}
		0 &= \rho\pdv[2]{u}{t} - \hat{k}\pdv[2]{u}{x} + \hat\gamma\pdv{u}{t} + \hat{v}(u^2 - \delta)(u - a) = 0,\label{eq:ContinuousModelDimensional}\\
		0 &= \pdv{a}{t} - \epsilon\left(\dfrac{u}{\eta} - a\right).\label{eq:ContinuousModelDimensional-A}
	\end{align}
A no flux boundary condition $\partial u / \partial x = 0$ is applied at $x = 0$ and $x = L$.

A key aspect of this study is to investigate the signal transmission speed through the parameter space, particularly as $\epsilon$ increases. We therefore non-dimensionalise \refeqs{eq:ContinuousModelDimensional}{eq:ContinuousModelDimensional-A} by scaling $t = T \hat{t}$, $x = X\hat{x}$, $u = U \hat{u}$ and $a = A \hat{a}$, where hat notation represents dimensionless variables. Choosing $U = A = \delta$, $T^2 = \rho / \hat{\gamma}$ and $X^2 = \hat{k} T^2 / \rho$, gives
	\begin{align}
		0 &= \pdv[2]{\hat{u}}{\hat{t}} - \pdv[2]{\hat{u}}{\hat{x}} + \pdv{\hat{u}}{\hat{t}} + \nu(\hat{u}^2-1)\left(\frac{\hat{u}}{\eta} -\hat{a}\right),\label{eq:ContinuousModel}\\
		0 &= \pdv{\hat{a}}{\hat{t}} - \kappa\left(\dfrac{\hat{u}}{\eta} - \hat{a}\right),\label{eq:ContinuousModel-A}
	\end{align}
where $\kappa = T\epsilon \ge 0$, $\nu = \hat{v}\delta^2\rho/\hat{\gamma}^2 > 0$, $\eta > 1$ and $\hat{x} \in (0,\hat{L})$. The behaviour of the system can now be studied through the three-dimensional parameter space $(\nu,\eta,\kappa)$ where: $\nu > 0$ is the relative strength of the potential function; $\eta > 1$ describes the steady state locations of $\hat{a}(\hat{x},\hat{t})$; and $\kappa \ge 0$ is the relative timescale of the reaction mechanism. In this non-dimensional regime, the stable states are located at $\hat{u} = \pm 1$ where the high potential energy state is always given by $\text{sign}(\hat{a})$. 

	\begin{figure}[!t]
		\centering
		\includegraphics[width=\textwidth]{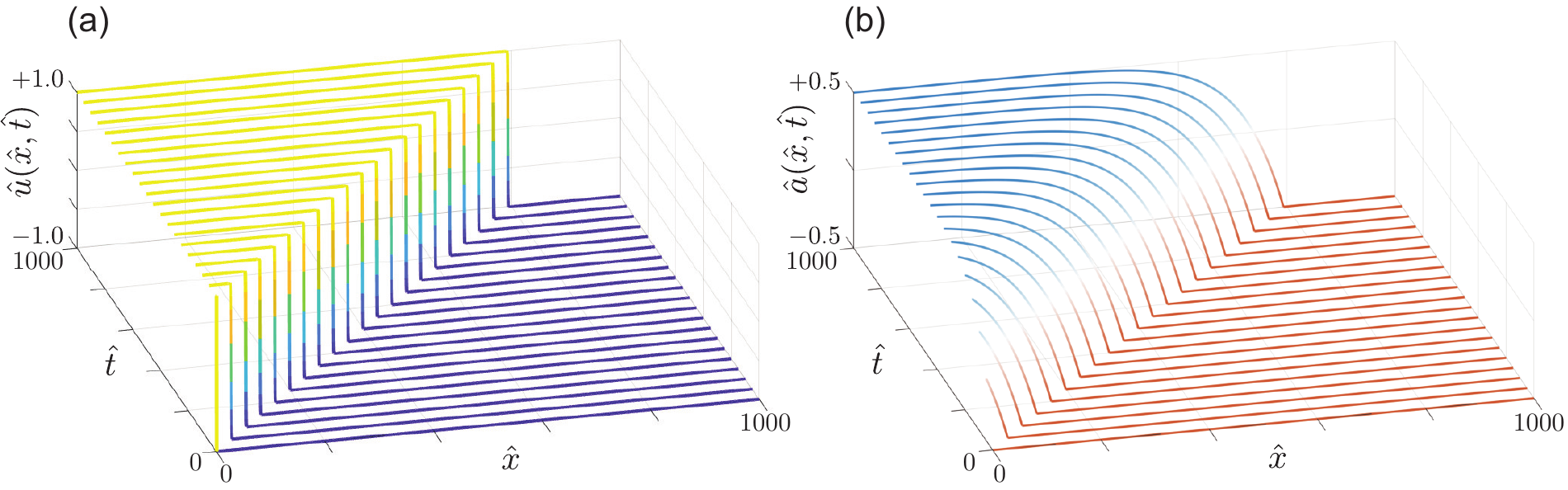}
		\caption{Numerical solutions to the non-dimensional continuous model (\refeqs{eq:ContinuousModel}{eq:ContinuousModel-A})  showing (a) the displacement field, $\hat{u}(\hat{x},\hat{t})$; and, (b) the biological reaction, $\hat{a}(\hat{x},\hat{t})$.  Parameters used are $\nu = 4$, $\eta = 2$ and $\kappa = 0.01$. The signal is initiated using the initial condition described by \refeqs{eq:ContinuousModelIC}{eq:ContinuousModelIC-A} with $Q = 1$.}
		\label{Fig-2-PDEWaveWaterfall}
	\end{figure}

Previous studies provide evidence to suggest that the transmitted energy is input-independent \cite{Hwang:2018jd}, so we do not expect the initial condition to affect the transmission speed in the centre of the domain. A signal is initiated in the continuous model using a Heaviside initial condition for the displacement where
	\begin{equation}\label{eq:ContinuousModelIC}
		\hat{u}(\hat{x},0) = \left\{\begin{array}{rl}
						1, & 0 \le \hat{x} < Q,\\
						-1, & Q \le \hat{x} \le \hat{L},
					\end{array}\right.
	\end{equation}
and the biological response is kept as it was before the signal was initiated by setting
	\begin{equation}\label{eq:ContinuousModelIC-A}
		\hat{a}(\hat{x},0) = -1/\eta.
	\end{equation}

In \reffig{Fig-2-PDEWaveWaterfall}{a} numerical solutions to the continuous model show that the transition of the displacement variable, $\hat{u}(\hat{x},\hat{t})$, is carried by a wave which appears to approach a constant shape and speed. \refFig{Fig-2-PDEWaveWaterfall}{b} demonstrates the slow biological response where $\hat{a}(\hat{x},\hat{t})$ also undergoes a slow transition in response to changes in $\hat{u}(\hat{x},\hat{t})$. Results in \reffig{Fig-3-Kymographs}{} illustrate the dependence of the transmission speed on the timescale of the response, $\kappa$. These results indicate a negative monotonic relationship between $\kappa$ and the transmission speed. Full details of the numerical scheme used to solve the continuous model are provided in the supporting material.

\section{Travelling Wave Model}
\label{s:TravellingWaveModel}

\refFig{Fig-2-PDEWaveWaterfall}{} suggests that signals are propagated through the system by waves which appear to approach a constant shape and speed. Motivated by this, we now look for a travelling wave solution to the continuous model by extending the domain to represent a material of infinite length, so that $\hat{x} \in \mathbb{R}^+$ \cite{Billingham:1991hu,Murray:2002uo,MathematicalPhysiologyI,Landman:2005fq,Simpson:2006il,Beck:2008bx,Rubin:2008kw,Nadkarni:2016jn,Nadkarni2016,Hwang:2018ju,Nizovtseva:2018gw}. We define the wavespeed, $c$, and without loss of generality enforce $c > 0$ by investigating travelling waves that are initiated at the left boundary and move in only the positive $\hat{x}$ direction. In reality, we expect symmetry in these solutions as a travelling wave may also travel in the negative $\hat{x}$ direction if the signal is initiated in the centre of the domain.

Substituting travelling wave variables $f(z) = \hat{u}(\hat{x},\hat{t})$ and $h(z) = \hat{a}(\hat{x},\hat{t})$, where $z = \hat{x} - c\hat{t} \in \mathbb{R}$, into \refeqs{eq:ContinuousModel}{eq:ContinuousModel-A}, and dividing by $c^2 - 1$, gives the travelling wave model:
	\begin{align}
		0 	&= \dv[2]{f}{z} + \dfrac{c}{1 - c^2}\dv{f}{z} - \dfrac{\nu}{1 - c^2}(f^2 - 1)(f - h),\:\:f(\pm\infty) = \mp 1,\label{eq:TravellingWaveModel}\\
		0 	&= \dv{h}{z} + \dfrac{\kappa}{c}\left(\dfrac{f}{\eta} - h\right),\:\:h(\infty) = -\dfrac{1}{\eta},\label{eq:TravellingWaveModel-H}
	\end{align}
which, in physical coordinates, corresponds to a wave profile connecting $\hat{u} = -1$ to $\hat{u} = 1$ as $t \rightarrow \infty$.
%
	\begin{figure}[!t]
		\centering
		\includegraphics[width=\textwidth]{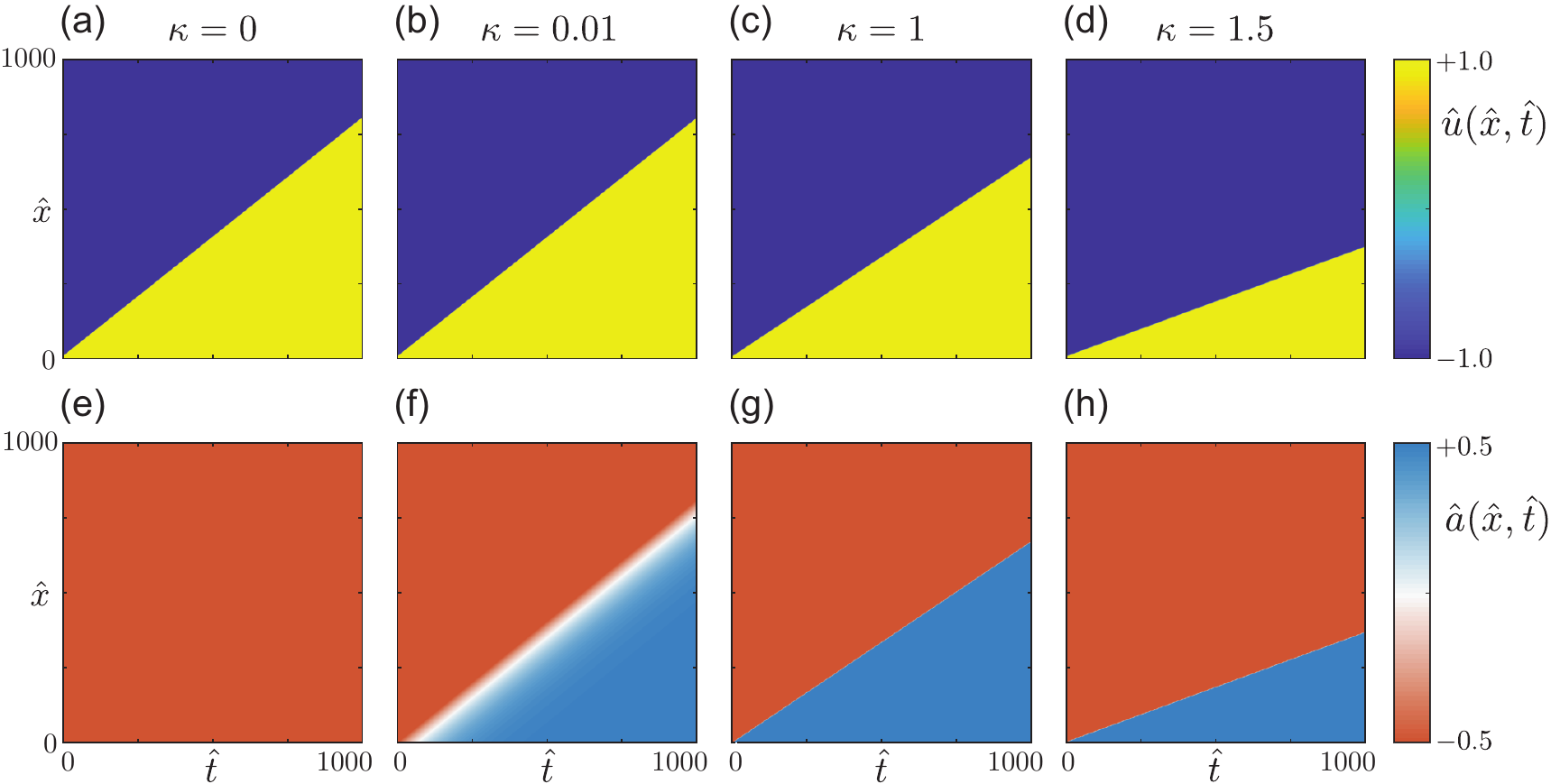}
		\caption{The solution to the continuous model, \refeq{eq:ContinuousModel}, shown as kymographs where colouring represents (a)-(d) the displacement function, $\hat{u}(\hat{x},\hat{t})$; and, (e)-(h) the biological response, $\hat{a}(\hat{x},\hat{t})$, for increasing values of $\kappa$. Parameter values used are $\nu = 4$ and $\eta = 2$. Each signal is initiated using the initial condition described by \refeqs{eq:ContinuousModelIC}{eq:ContinuousModelIC-A} with $Q = 1$.}
		\label{Fig-3-Kymographs}
	\end{figure}

In \reffig{Fig-3-Kymographs}{} numerical solutions of the continuous model show that increasing the response speed $\kappa$ typically reduces the wavespeed from a maximum which occurs when $\kappa = 0$. We therefore denote a \textit{fast wave} as a travelling wave solution at, or near, $\kappa = 0$, and a \textit{slow wave} as a travelling wave solution in the limit, or near, $c = 0$. Removing the active component of the system from the model by setting $\kappa = 0$ (and defining $c_0$ as the corresponding wavespeed) gives:

	\begin{align}
		0 	&= \dv[2]{f_0}{z} + \dfrac{c_0}{1 - c_0^2}\dv{f_0}{z} - \dfrac{\nu}{1 - c_0^2}(f_0^2 - 1)\left(f_0+ \dfrac{1}{\eta}\right),\:f_0(\pm\infty) = \mp 1,\label{eq:TravellingWaveModel_Fast}\\
		0 	&= \dv{h_0}{z},\:\:h_0(\infty) = -\dfrac{1}{\eta},\label{eq:TravellingWaveModel_Fast_H}
	\end{align}
where $f_0(z)$ will correspond to the solution to the fast wave. Under certain parameter transformations, \refeq{eq:TravellingWaveModel_Fast} is analogous to the well-studied bistable equation, that arises from analysis of the FitzHugh-Nagumo model \cite{FitzHugh1955,MathematicalPhysiologyI,Beck:2008bx,Rubin:2008kw}.  The solution of \refeqs{eq:TravellingWaveModel_Fast}{eq:TravellingWaveModel_Fast_H} is therefore
	\begin{align}
		f_0(z) &= -\tanh\left(\mu_0z\right),\label{eq:TravellingWaveModel_Fast_Solution}\\
		h_0(z) &= -\dfrac{1}{\eta},
	\end{align}

where
	\begin{equation}\label{eq:TravellingWaveModel_Fast_C0Mu0}
		\begin{array}{ccc}
			\mu_0 = \sqrt{\dfrac{\nu^2}{\eta^2} + \dfrac{\nu}{2}} &\text{ and}& c_0 = \sqrt{\dfrac{1}{\dfrac{\eta^2}{2\nu}+1}}.
		\end{array}
	\end{equation}
For physically realistic parameter combinations, \refeq{eq:TravellingWaveModel_Fast_C0Mu0} suggests that $c_0 < 1$.

%
\begin{figure}[!t]
	\centering
	\includegraphics[width=\textwidth]{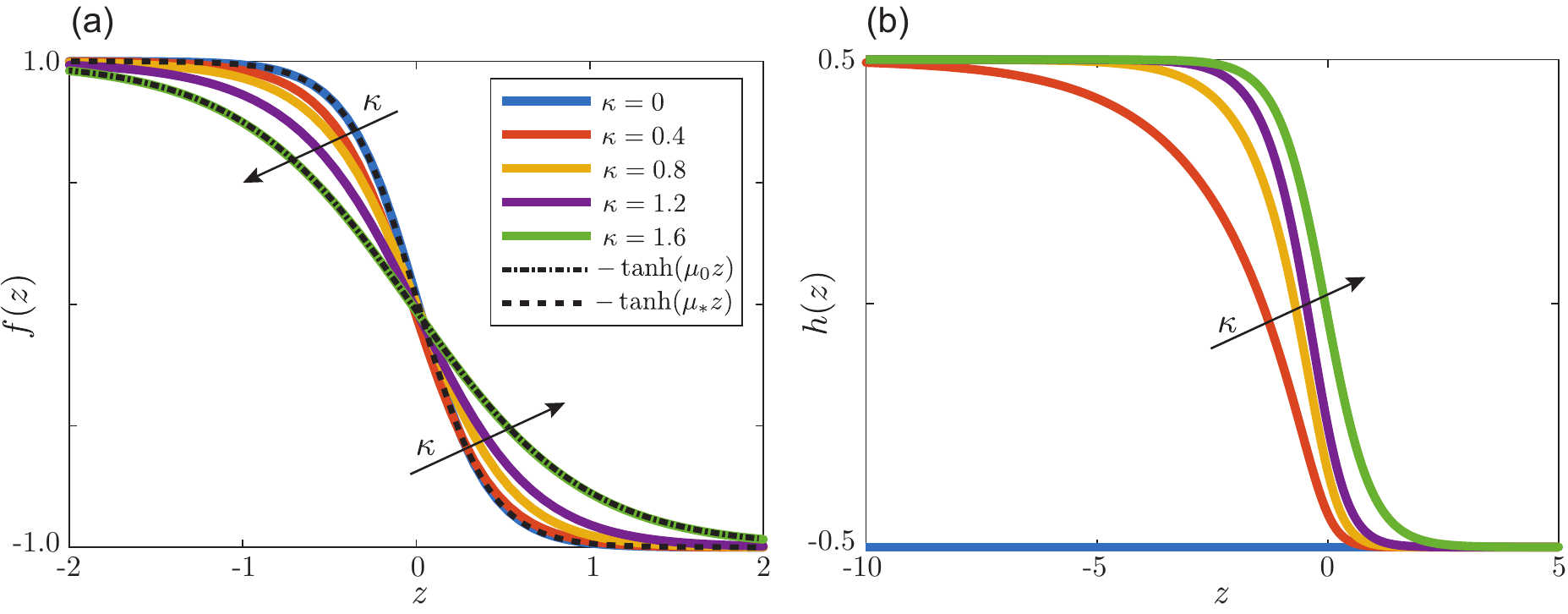}
	\caption{ Approximate numerical solutions to the travelling wave model, formulated from numerical solutions to the continuous model, for (a) the displacement function, $f(z)$; and, (b) the biological reaction, $h(z)$. Results are shown for various values of the response parameter, $\kappa$, with $\nu = 4$ and $\eta = 2$. Additionally, (a) shows the exact solutions to the travelling wave model, \refeqs{eq:TravellingWaveModel}{eq:TravellingWaveModel-H}, for $\kappa = 0$ (black dashed-dot) and $\kappa = \kappa_* = 1.6$ (black dashed).}
	\label{Fig-4-WaveShape}
\end{figure}
%

\subsection{Perturbation solution for the fast wave}
\label{ss:PerturbationSolution}

\begin{figure}[!t]
		\centering
		\includegraphics[width=\textwidth]{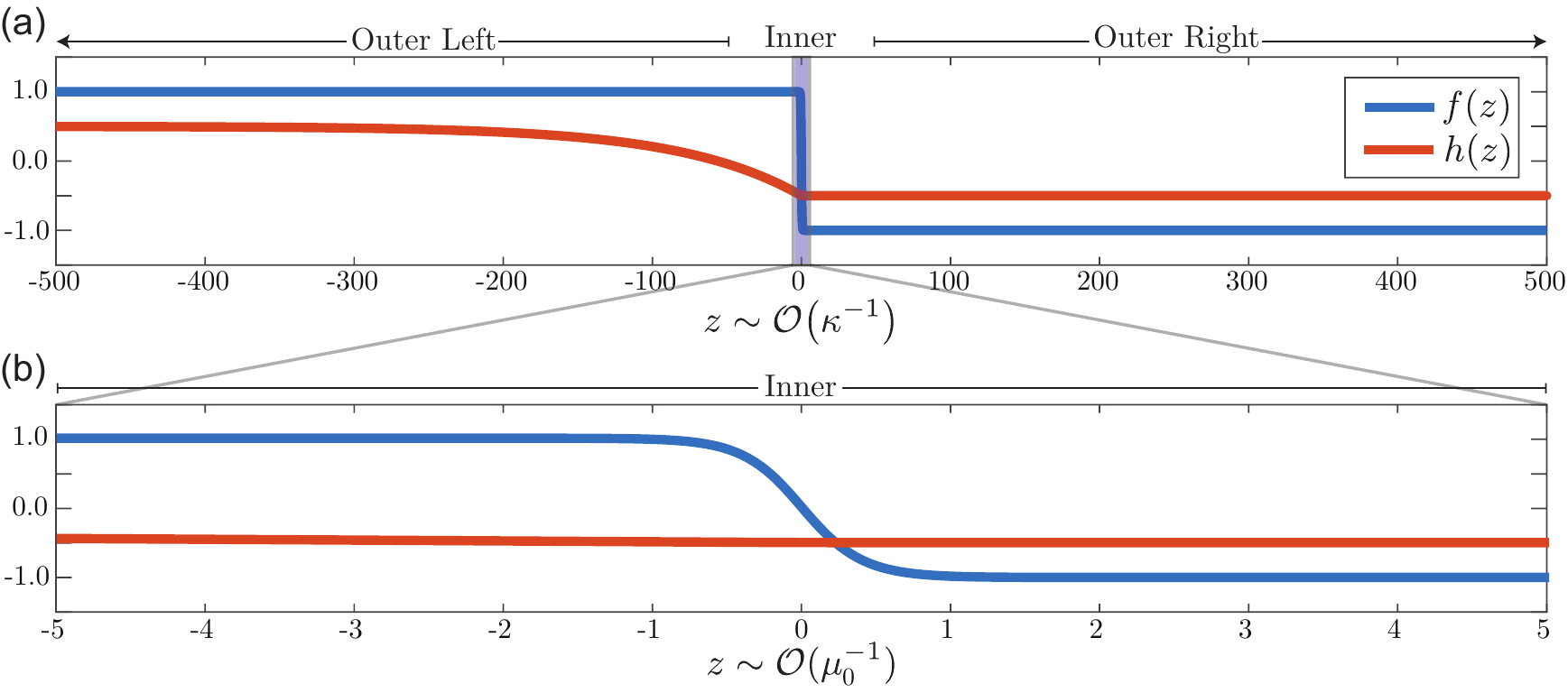}
		\caption{ Approximation of the travelling wave profile using a solution to the continuous model for (a) the displacement function, $f(z)$; and, (b) the biological reaction, $h(z)$.  Parameters used are $\nu = 4$, $\eta = 2$ and $\kappa = 0.01$. In (a) the horizontal axis is scaled to show both outer regions where $z \sim \bigok{-1}$, which is the rate of change of $h(z)$. In (b) the horizontal axis is scaled to show the inner region where $z \sim \mathcal{O}(\mu_0^{-1})$, which is the rate of change of $f(z)$.}
		\label{Fig-5-BoundaryLayer}	
	\end{figure}

An exact solution in the limiting case $\kappa = 0$ (\refeq{eq:TravellingWaveModel_Fast_Solution}) allows the formulation of a perturbation solution for $0 < \kappa \ll 1$ \cite{Hinch1991,MathematicalPhysiologyI,Landman:2005fq}. \refFig{Fig-5-BoundaryLayer}{} shows an approximation to the wave profiles $f(z)$ and $h(z)$ for $\kappa = 0.01$. A fast transition region is seen in $f(z)$ around $ z= 0$ (\reffig{Fig-5-BoundaryLayer}{a}) suggesting behaviour necessitating a singular perturbation analysis \cite{Hinch1991,Landman:2005fq}. For $z \sim \mathcal{O}(\mu_0^{-1})$ the solution appears to match a solution for $\kappa = 0$, since, at this scale, $h(z)$ is approximately constant (\reffig{Fig-5-BoundaryLayer}{b}). In \reffig{Fig-5-BoundaryLayer}{a} we show that $h(z)$ is not constant but rather a slow reaction of $\mathcal{O}(\kappa^{-1})$ which transitions $h(z) = -{1}/{\eta}$ to $h(z) = {1}/{\eta}$ as $z \rightarrow -\infty$. This observation further suggests a singular perturbation analysis, since the behaviour of the response mechanism $h(z)$ varies significantly for $0 < \kappa \ll 1$ compared to $\kappa = 0$.

%

We propose a three-part perturbation solution about $\kappa = 0$ and define independent variables $z~\sim~\mathcal{O}(\mu_0^{-1})$ to correspond to an \textit{inner region}, and $Z = \kappa z$ for $z \sim \mathcal{O}(\kappa^{-1})$, to correspond to two \textit{outer regions}. These three regions are shown in \reffig{Fig-5-BoundaryLayer}{a}. This regime requires the fast process, which occurs in the inner region, to be much faster than the slow process, which occurs in the outer region. That is, we require $\kappa \ll \mu_0$. 

Our aim in looking for a perturbation solution is to determine the effect of small perturbations in $\kappa$ on $c$. To do this, we pose a perturbation expansion for the wavespeed through the entire domain,
	\begin{equation}\label{eq:PertubationExpansion_C}
		c = c_0 + c_1\kappa + \bigok{2},
	\end{equation}
where $c_0$ is given by \refeq{eq:TravellingWaveModel_Fast_C0Mu0}.

\begin{figure}[!t]
		\centering
		\includegraphics[width=\textwidth]{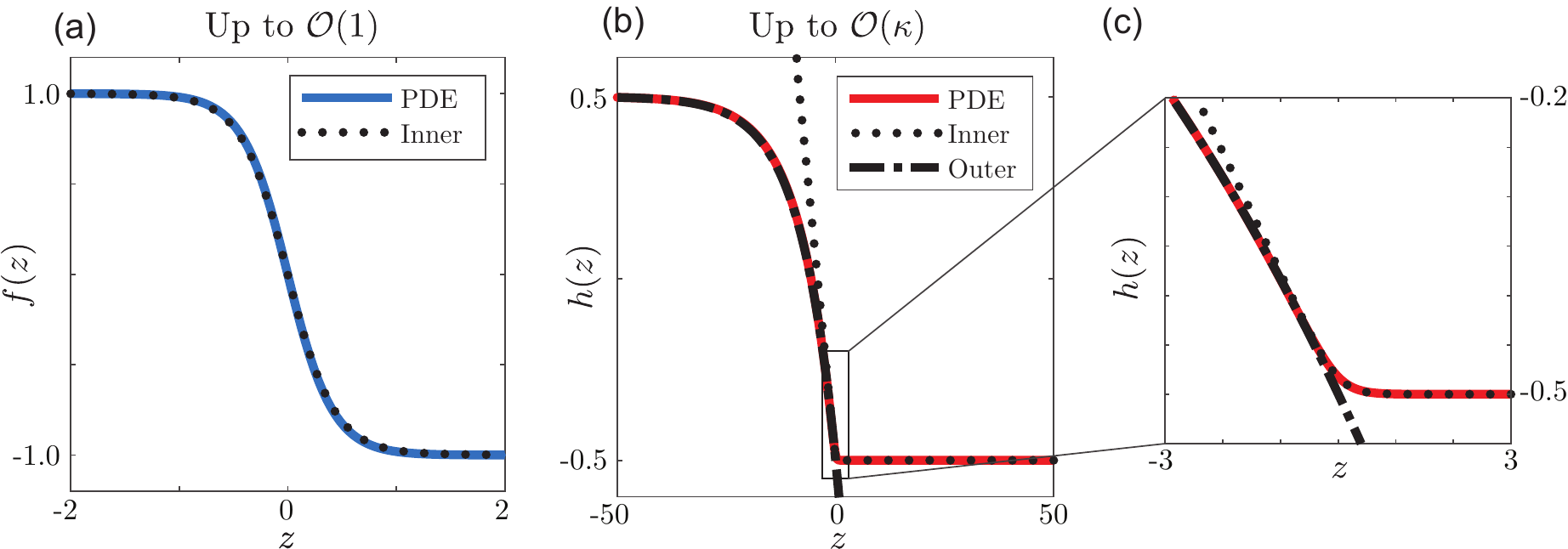}
		\caption{Comparison of the wave shape from a numerical solution to continuous model (coloured solid curves), to the perturbation solution (black dotted and dotted-dashed curves) for (a) the displacement function, $f(z)$, to $\mathcal{O}(1)$ and (b) the reaction mechanism, $h(z)$, to $\mathcal{O}(\kappa)$. The inset (c) shows the crossover between the inner and outer perturbation solutions to $h(z)$ in further detail, showing an excellent match to the wave shape from the continuous model, and between inner and outer solutions.}
		\label{Fig-6-PerturbationSolutions}
	\end{figure}

In the inner region, we pose a perturbation solution of the form
	\begin{align}
		f(z) &= f_0(z) + f_1(z) \kappa + \bigok{2},\\
		h(z) &= h_0(z) + h_1(z) \kappa + \bigok{2},
	\end{align}
where $f_0(z)$ and $h_0(z)$ correspond to the shape of the fast wave at $\kappa = 0$, given by \refeq{eq:TravellingWaveModel_Fast_Solution}. Full details of the perturbation solution in the inner region are given in the supporting material. In summary, the solution is given by
	\begin{align}
		f(z) &= -\tanh(\mu_0 z) + f_1(z)\kappa + \bigok{2},\label{eq:S-TW-FastSolution-f}\\
		h(z) &= -\dfrac{1}{\eta} + \dfrac{1}{c_0 \mu_0 \eta} \bigg(\log(\cosh(\mu_0 z)) - \mu_0 z + \log(2)\bigg)\kappa + \bigok{2},\label{eq:S-TW-FastSolution-h}
	\end{align}
where $f_1(z)$ and $c_1$ are defined by the solution of a second-order boundary value problem, which can be solved numerically. Full details of this numerical scheme are given in the supporting material. These solutions are shown in \reffig{Fig-6-PerturbationSolutions}{}, and the wavespeed correction, $c_1$, is summarised for various parameter combinations in table S3 in the supporting material.
%

In the outer region we denote solutions to the slow system using uppercase variables, $F(Z)$ and $H(Z)$. We expect the outer solution to apply for $Z \sim \mathcal{O}(1) \Rightarrow z \sim \mathcal{O}(\kappa^{-1})$ and as $z,Z \rightarrow \infty$. In the outer region, \refeq{eq:TravellingWaveModel} becomes
	\begin{equation}\label{eq:PerturbationExpansion_SlowModel}
		0 	= \kappa^2\dv[2]{F}{Z} + \dfrac{c\kappa}{1- c^2}\dv{F}{Z}- \dfrac{\nu}{1- c^2}(F^2-1)(F-H),\:\:F(\pm\infty) = \mp 1,
	\end{equation}
which has the solution
	\begin{equation}\label{eq:PerturbationExpansion_SlowModel_Solution-F}
		F(Z) = -\text{sign}(Z).
	\end{equation}
This agrees with the sharp transition region seen at this scale in \reffig{Fig-5-BoundaryLayer}{a}. Substituting \refeq{eq:PerturbationExpansion_SlowModel_Solution-F} into \refeq{eq:TravellingWaveModel-H} we see that
	\begin{equation}\label{eq:S-TW-SlowODE}
		0 	= \dv{H}{Z} - \dfrac{1}{c}\left(\dfrac{\text{sign}(Z)}{\eta} + H\right),\:\: H(\infty) = -\frac{1}{\eta}.\\
	\end{equation}
Full details of perturbation solution in the outer region are given in the supporting material. In summary, the solution to \refeq{eq:S-TW-SlowODE} is given by
	\begin{equation}\label{eq:S-TW-SlowSolution-H}
		H(Z) = \left\{\begin{array}{ll}
					 \dfrac{1}{\eta} - \dfrac{2}{\eta} \exp\left(\dfrac{Z}{c_0}\right) - \dfrac{2c_1}{\eta c_0^2}Z\exp\left(\dfrac{Z}{c_0}\right)\kappa + \bigok{2}, & Z < 0,\\
					-\dfrac{1}{\eta}, & Z > 0.
			\end{array}\right.
	\end{equation}

\refFig{Fig-6-PerturbationSolutions}{b-c} shows a comparison between solutions for the reaction mechanism in the inner and outer regions (given by \refeqs{eq:S-TW-FastSolution-f}{eq:S-TW-SlowSolution-H}, respectively), to an approximation of the wave shape formed form the numerical solution to the continuous model. We see an excellent match between both the continuous model and the perturbation solutions, as well as between the inner and outer regions of the perturbation solution. These results provide excellent information about the behaviour of $h(z)$ for $0 < \kappa \ll 1$, and are particularly important as the behaviour for $0 < \kappa \ll 1$ varies significantly from the behaviour at $\kappa = 0$.

\subsection{Energy transport}
\label{ss:EnergyTransport}

To determine information about the slow wave, and to obtain more information about the fast wave,  we follow Nadkarni \etal \cite{Nadkarni:2016jn,Nadkarni2016} to derive an integrability condition -- that is, a necessary condition for existence of a solution with given boundary conditions -- to investigate the transported energy.   Multiplying the travelling wave model, given by \refeq{eq:TravellingWaveModel},  by $\text{d}f/\text{d}z$ and integrating gives
	\begin{equation}\label{eq:TW_Integrated}
		0 = \intinf \dv[2]{f}{z}\dv{f}{z} \dz + \dfrac{c}{1 - c^2} \intinf \left(\dv{f}{z}\right)^2 \dz - \dfrac{\nu}{1 - c^2} \intinf (f^2 - 1)(f - h)\dv{f}{z} \dz.
	\end{equation}

For a parameter regime where the transition wave exists, the velocity will vanish in the far field so that $f \rightarrow \mp 1$ and $\text{d}f/\text{d}z \rightarrow 0$ as $z \rightarrow \pm \infty$. Therefore some components of \refeq{eq:TW_Integrated} vanish:
	\begin{align*}
		\intinf \dv[2]{f}{z}\dv{f}{z} \dz &= \left[\dfrac{1}{2}\left(\dv{f}{z}\right)^2\right]_{-\infty}^\infty = 0,\\
		\text{and }\intinf (f^2 - 1)(f - h)\dv{f}{z} \dz &= -\intinf \dv{f}{z} (f^2 - 1)h \dz.
	\end{align*}
Under the assumption $|c| \neq 1$, \refeq{eq:TW_Integrated} becomes an \textit{integrability condition},
	\begin{equation}\label{eq:Integrability-Condition}
		c\intinf \left(\dv{f}{z}\right)^2 \dz = -\nu\intinf \dv{f}{z} (f^2 - 1) h \dz.
	\end{equation}
It is useful to note that, for $f = -\tanh(\sigma z)$, for some $\sigma$, which occurs as $\kappa \rightarrow 0$ with $\sigma = \mu_0$,
	\begin{equation}
		(f^2 - 1)\dv{f}{z} h \propto \sech^4(\sigma z) h.
	\end{equation}
This suggests that only the component of $h(z)$ near $z = 0$ is important in gaining any approximation from the integrability condition, provided $f(z)$ has the form of a hyperbolic tangent function. Since a travelling wave connecting $f(z)~=~1$ to $f(z)~=~-1$ as $z \rightarrow \infty$ will always have a sigmoidal form (\reffig{Fig-4-WaveShape}{}), we expect this observation to apply for all regions of the parameter space where a travelling wave exists.

In the inactive model, where $\kappa = 0$, Nadkarni \etal \cite{Nadkarni2016} show that the integrability condition reduces to
	\begin{equation}
		E_k = \dfrac{c \Delta V}{2\gamma},
	\end{equation}
where $E_k$ represents the total kinetic energy per density transported by the transition wave, and $\Delta V$ represents an \textit{energy gap} or difference the potential energy between the high and the low potential energy states. This result can be used to find an upper bound for $c$ for $\kappa~>~0$: Since $|h(t)|~\le~1/\eta$, and $\Delta V$ is a monotonically decreasing function of $h(t)$, the available kinetic energy in the system is always bounded above by that which occurs when $h(t) \equiv 1/\eta$, which occurs for $\kappa~=~0$. This suggests that $c(\kappa) \le c_0 < 1$ and substantiates numerical evidence seen in \reffig{Fig-3-Kymographs}{} which suggests a decreasing monotonic relationship between $c$ and $\kappa$. 

In the following subsections we apply the integrability condition to obtain approximations to $c(\kappa)$ while holding $\nu$ and $\eta$ constant. We also find the region of the parameter space that allows signal transmission. The advantage of these approximations is that they avoid numerical solutions to the continuous model to approximate the wavespeed. Numerical solutions to this model are not computationally inexpensive and it is generally difficult to obtain and verify the results.

\subsubsection{Energy transport in the fast wave}
\label{sss:EnergyTransportFast}

For $\kappa \ll 1$ it is reasonable to assume that $f \approx -\tanh(\mu z)$ where $\mu$ depends on $\kappa$, and $\mu \rightarrow \mu_0$ as $\kappa \rightarrow 0$. By assuming $f(z)$ has a similar form to $f_0(z)$ for $\kappa \ll 1$, it is reasonable to use the perturbation solution (\refeq{eq:S-TW-FastSolution-h}) as an approximation for $h(z)$. Since the integrability condition (\refeq{eq:Integrability-Condition}) depends only on the component of $h(z)$ near $z = 0$, we use only \refeq{eq:S-TW-FastSolution-h}. Allowing $\mu_0 = \mu(0)$ and $c_0 = c(0)$ where $\mu = \mu(\kappa)$ and $c = c(\kappa)$ depend on $\kappa$, we assume that
	\begin{equation}\label{eq:TravellingWaveModel_Fast_Hz}
		h(z) \approx -\dfrac{1}{\eta} + \dfrac{\kappa}{c\mu\eta} \bigg(\log(\cosh(\mu z)) - \mu z + \log(2)\bigg),\:\:\kappa \ll 1.
	\end{equation}

\refEq{eq:TravellingWaveModel_Fast_Hz} corresponds to the \textit{smoothed} piecewise-defined function where growth is equal to $2\kappa/(c\eta)$ for $z < 0$, and zero for $z > 0$. This function corresponds exactly to an $\mathcal{O}(1)$ approximation to $h(z)$ which uses $f(z) = -\text{sign}(z)$. Substituting \refeq{eq:TravellingWaveModel_Fast_Hz} into the integrability condition (\refeq{eq:Integrability-Condition}) provides the relationship
	\begin{equation}\label{eq:TravellingWaveModel_Fast_Eq}
		\frac{4c^2\mu^2}{3} = \frac{2\nu(6c\mu - 5\kappa)}{9\eta}.
	\end{equation}
This result does not allow $c$ and $\mu$ to be determined independently, so we consider a far field expansion of the travelling wave \cite{Billingham:1991hu}, that is, an expansion about $\phi \approx 0$, where $\phi = e^{-\mu z}$:
	\begin{equation}\label{eq:FarField_fz}
		f(z) = -\tanh(\mu z) = \dfrac{e^{-2\mu z} - 1}{e^{-2\mu z} + 1} = \dfrac{\phi^2 - 1}{\phi^2 + 1} = -1 + 2\phi^2 + \mathcal{O}(\phi^4).
	\end{equation}
Substituting $\phi = e^{-\mu z}$ and \refeq{eq:FarField_fz} into the travelling wave equation for $h(z)$ (\refeq{eq:TravellingWaveModel-H}) gives
	\begin{equation}\label{eq:FarFieldExpansions}
		\begin{aligned}
			f(z) &\sim -1 + 2e^{-2\mu z},\:\:|z^{-1}| \ll 1,\\
			h(z) &\sim -\dfrac{1}{\eta} + \dfrac{2\kappa}{\eta(\kappa + 2c\mu)}e^{-2\mu z},\:\:|z^{-1}| \ll 1.\\
		\end{aligned}
	\end{equation}
Substituting \refeq{eq:FarFieldExpansions} into \refeq{eq:TravellingWaveModel} provides a far-field relationship,
	\begin{equation}\label{eq:TravellingWaveModel_FarFieldEq}
		0 = \frac{\nu}{\eta}-c\mu-\nu+2\mu^2(1-c^2).
	\end{equation}

\begin{figure}[!t]
	\centering
	\includegraphics[width=\textwidth]{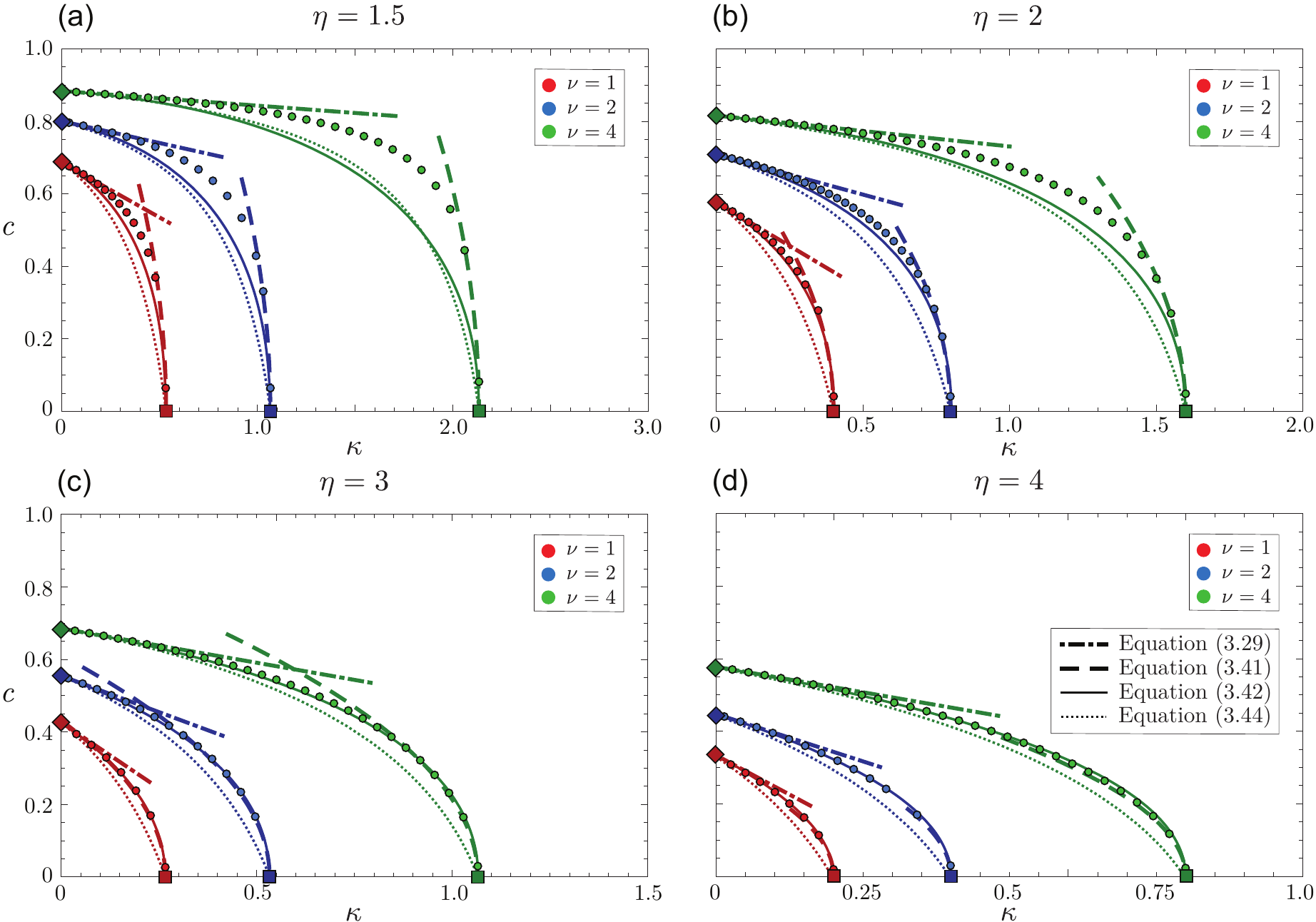}
	\caption{Comparison between the wavespeed as a function of $\kappa$, from the continuous model (circles) to: the analytical expression at $\kappa = 0$, given by \refeq{eq:TravellingWaveModel_Fast_C0Mu0} (diamonds); the analytical expression for $\kappa_*$ at $c = 0$, given by \refeq{eq:kappa_star} (squares); the analytical approximation given by \refeq{eq:Approximation-FastWave}, which applies for $\kappa \ll 1$ (dashed-dotted curve); the analytical approximation given by \refeq{eq:IntegrabilityEq_Slow_Curve}, which applies for $c \ll 1$ (dashed-curve); the combined approximation given by \refeq{eq:CombinedForm} which applies everywhere (dotted curve); and, the approximation given by \refeq{eq:WholeAnsatz} which applies everywhere (solid curve).}
	\label{Fig-7-Approximations}
\end{figure}

To obtain a useful analytical expression for $c$ and $\mu$, we pose a perturbation solution to  \refeqs{eq:TravellingWaveModel_Fast_Eq}{eq:TravellingWaveModel_FarFieldEq} around $\kappa = 0$, such that
	\begin{align}
		c 	&= c_0 + \tilde{c}_1\kappa + \bigok{2},\\
		\mu &= \mu_0 + \tilde{\mu}_1\kappa + \bigok{2},
	\end{align}
where $c_0$ and $\mu_0$ are given by \refeq{eq:TravellingWaveModel_Fast_C0Mu0}. We note that $\tilde{c}_1$ approximates $c_1$, the gradient of $c(\kappa)$ at $\kappa = 0$, where $c_1$ is defined exactly by the solution of the perturbation problem. Substitution into \refeqs{eq:TravellingWaveModel_Fast_Eq}{eq:TravellingWaveModel_FarFieldEq} gives
	\begin{equation}\label{eq:c1_approximate}
		\tilde{c}_1 = \frac{5 \nu  \left(4 c_0^2 \mu _0+c_0-4 \mu _0\right)}{24 \mu_0^2 \left(2 c_0 \eta  \mu _0-\nu \right)}.
	\end{equation}
We compare this estimate of $c_1$ from that calculated numerically by solving the boundary value problem that comes from the singular perturbation expansion, and that estimated from the continuous model, in table S3 in the supporting material. In \reffig{Fig-7-Approximations}, we show that, for $\kappa \ll 1$,
	\begin{equation}\label{eq:Approximation-FastWave}
		c(\kappa) \sim c_0 + \tilde{c}_1\kappa,
	\end{equation}
matches numerical results for $c(\kappa)$.

%
%

\subsubsection{Energy transport in the slow wave}
\label{sss:EnergyTransportSlow}

We denote $f_*(z)$ and $h_*(z)$ as the shape of the slow wave, which occurs as $c \rightarrow 0$, and $\kappa \rightarrow \kappa_*$, where $\kappa_*$ is yet to be determined. In addition, we expect the curve $c(\kappa)$ to be perpendicular to the $\kappa$ axis at $\kappa_*$ to maintain continuity in the symmetry of the problem where solutions with a negative wavespeed are equally valid. To determine a governing equation for the slow wave, we take $c \rightarrow 0$, so that \refeqs{eq:TravellingWaveModel}{eq:TravellingWaveModel-H} become
	\begin{align}\label{eq:TravellingWaveModel_Slow}
		0 &= \dv[2]{f_*}{z} - \nu \left(1 - \dfrac{1}{\eta}\right)f_*(f_*^2 - 1),\:\: f_*(\pm \infty) = \mp 1,\\
		0 &= \dfrac{f_*}{\eta} - h_*,
	\end{align}
which have the analytical solution
	\begin{equation}\label{eq:S-C0_Solutions}
		\begin{aligned}
			f_*(z) &= -\tanh\left(\mu_*z\right),\\
			h_*(z) &= -\dfrac{1}{\eta}\tanh\left(\mu_*z\right),
		\end{aligned}
	\end{equation}
where
	\begin{equation}\label{eq:TravellingWaveModel_MuStar}
		\mu_* = \sqrt{\dfrac{\nu}{2}\left(1 - \dfrac{1}{\eta}\right)}.
	\end{equation}

Direct substitution of $f_*(z)$ and $h_*(z)$ into the integrability condition (\refeq{eq:Integrability-Condition}) causes all terms to vanish, so higher-order behaviour of $\mathcal{O}(c)$ as $c \rightarrow 0$ is important. To allow for this, we note that \refeq{eq:TravellingWaveModel-H} can be solved, for all $\kappa$, to give
	\begin{equation}\label{eq:TravellingWaveModel_HIntegralSolution}
		h(z) = \dfrac{\kappa}{c\eta}\int_z^\infty f(s) \exp\left[\dfrac{\kappa}{c}(z - s)\right]\:\text{d}s.
	\end{equation}
\refEq{eq:TravellingWaveModel_HIntegralSolution} shows that $h(z)$ depends the product of $f(s)$ and a function that decays rapidly as $z$ moves away from $s$. Expanding $f(s)$ in a Taylor series about $z$ gives
	\begin{align}\label{eq:TravellingWaveModel_HIntegralSolution_Series}
		h(z) &\approx \dfrac{\kappa}{c\eta}\int_z^\infty\left[\sum_{n=0}^\infty \dfrac{(s-z)^n}{n!}\dv[n]{f}{z}\right]\exp\left(\dfrac{\kappa}{c}(z - s)\right)\:\text{d}s
			= \dfrac{1}{\eta}\sum_{n=0}^\infty \dfrac{c^n}{\kappa^n}\dv[n]{f}{z}.
	\end{align}

Assuming that $f \sim \tanh(\mu(\kappa) z)$ provided $c \ll 1$, where $\mu(\kappa_*) = \mu_*$, and truncating the infinite series given by \refeq{eq:TravellingWaveModel_HIntegralSolution_Series} after $n = 3$, the integrability condition (\refeq{eq:Integrability-Condition}) gives the relationship
	\begin{equation}\label{eq:IntegrabilityEq_Slow}
		\frac{4\kappa^3}{3}=\frac{16\nu\left(7 \kappa ^2 - 8 c^2 \mu ^2\right)}{105\eta}.
	\end{equation}
For $\eta$ and $\nu$ fixed, $c \rightarrow 0$ only as $\kappa \rightarrow \kappa_*$. Substituting $c = 0$ into \refeq{eq:IntegrabilityEq_Slow} gives
	\begin{equation}\label{eq:kappa_star}
		\kappa_* = \dfrac{4}{5}\dfrac{\nu}{\eta}.
	\end{equation}
Under the assumption that $c(\kappa)$ is monotonically decreasing, the result in \refeq{eq:kappa_star} provides an analytical expression for the region of the parameter space where we expect signal transmission. In \reffig{Fig-7-Approximations}{} we show that this expression matches the numerical results. Interestingly, \refeq{eq:kappa_star} depends only on the ratio of the other parameters, $\nu/\eta$. The scales of the horizontal axes for figure pairs \reffig{Fig-7-Approximations}{a,c} and \reffig{Fig-7-Approximations}{b,d} have been chosen to highlight this. 

To gain information about the shape of $c(\kappa)$ near $\kappa = 0$, we pose a perturbation solution around $\kappa = \kappa_*$ to the system governed by \refeq{eq:IntegrabilityEq_Slow} and the far-field relationship, \refeq{eq:TravellingWaveModel_FarFieldEq}, where
	\begin{align}
		c 	&= c_* + c_\frac{1}{2}\sqrt{\kappa_*-\kappa} + \mathcal{O}\left(\kappa_*-\kappa\right),\\
		\mu &= \mu_* + \mu_\frac{1}{2}\sqrt{\kappa_*-\kappa} + \mathcal{O}\left(\kappa_*-\kappa\right),
	\end{align}
where $\mu_*$, given by \refeq{eq:TravellingWaveModel_MuStar}, and $c_* = 0$ apply at $\kappa = \kappa_*$. Substitution of these expansions into \refeqs{eq:TravellingWaveModel_FarFieldEq}{eq:IntegrabilityEq_Slow} gives
	\begin{equation}\label{eq:IntegrabilityEq_Slow_CHalf}
		c_\frac{1}{2} = \sqrt{\dfrac{7}{5(\eta - 1)}},
	\end{equation}
so that, for $\kappa \approx \kappa_*$,
	\begin{equation}\label{eq:IntegrabilityEq_Slow_Curve}
		c(\kappa) \sim \sqrt{\dfrac{7(\kappa_* - \kappa)}{5(\eta - 1)}}.
	\end{equation}
\refFig{Fig-7-Approximations}{} shows that \refeq{eq:IntegrabilityEq_Slow_Curve} matches numerical results for $c(\kappa)$ for a surprisingly wide range of $\kappa < \kappa_*$, especially for larger $\eta$. This result is particularly important as we have not constructed a full perturbation solution to the travelling wave model about $c = 0$.

\subsubsection{Combined approximation}
\label{sss:WholeDomain}

We can combine the approximations to $c(\kappa)$ from the slow and the fast wave to obtain a curve that behaves like \refeq{eq:Approximation-FastWave} for $\kappa \rightarrow 0$ and like \refeq{eq:IntegrabilityEq_Slow_Curve} as $\kappa \rightarrow \kappa_*$. To do this, we propose a form like
	\begin{equation}\label{eq:CombinedForm}
		c(\kappa) \approx \alpha_1 + \alpha_2 \kappa + \alpha_3\sqrt{\kappa_* - \kappa},
	\end{equation}
where we choose
	\begin{equation}
	\begin{aligned}
		\alpha_1 &= -(c_0 + 2\tilde{c}_1\kappa_*), &
		\alpha_2 &= \dfrac{c_0}{\kappa_*} + 2\tilde{c}_1, &
		\alpha_3 &= \dfrac{2(c_0 + \tilde{c}_1\kappa_*)}{\sqrt{\kappa_*}}	
	\end{aligned}
	\end{equation}
where $c_0$ is given by \refeq{eq:TravellingWaveModel_Fast_C0Mu0}; $\tilde{c}_1$ is given by \refeq{eq:c1_approximate}; and, $\kappa_*$ is given by \refeq{eq:kappa_star}. In \reffig{Fig-7-Approximations}{} we show that this combined approximation provides a reasonable approximation to $c(\kappa)$ for $\kappa \in [0,\kappa_*]$.

\subsubsection{Whole domain ansatz}
\label{sss:WholeDomain}

 Results in \refeqs{eq:TravellingWaveModel_Fast_Solution}{eq:S-C0_Solutions} show that $f(z)$ is described exactly by a hyperbolic tangent at both $\kappa = 0$ and $\kappa = \kappa_*$, and \reffig{Fig-4-WaveShape}{a} suggests that $f(z)$ remains sigmoidal.  Therefore, it may be reasonable to approximate $f(z) \approx \tanh(\mu(\kappa) z)$ for $\kappa \in [0,\kappa_*]$, where $\mu(\kappa)$ depends on $\kappa$. Additionally assuming that $\mu(\kappa)$ decays linearly from $\mu = \mu_0$ at $\kappa = 0$ to $\mu = \mu_*$ at $\kappa = \kappa_*$, we obtain an approximation to $f(z)$ for all $\kappa$:
	\begin{equation}\label{eq:WholeAnsatz}
		f(z) \sim -\tanh\left(\mu(\kappa) z \right),\:\:\mu(\kappa) = \dfrac{\mu_* - \mu_0}{\kappa_*}\kappa+ \mu_0.
	\end{equation}
Substituting the approximation for $\mu(\kappa)$ given by \refeq{eq:WholeAnsatz} into the far-field matching condition (\refeq{eq:TravellingWaveModel_FarFieldEq}) gives an approximation to $c(\kappa)$ which can apply for $\kappa \in [0,\kappa_*]$. We show that this approximation is reasonably accurate throughout $\kappa \in [0,\kappa_*]$ in \reffig{Fig-7-Approximations}{},  however it is clear from numerical results in \reffig{Fig-4-WaveShape}{}, when $\kappa = 0.4$, that the solution does not have the symmetry of a hyperbolic tangent function.

\section{Discussion and Conclusion}
\label{s:Discussion}

Currently, the inactive metamaterial described mathematically by Nadkarni \etal \cite{Nadkarni2016} and experimentally realised by Raney \etal \cite{Raney2016} is able to transmit mechanical signals by the release of stored potential energy. A limitation of this design is that mechanical energy must be manually introduced into the system before additional signals can be transmitted. Our study presents a novel biologically inspired metamaterial that incorporates a theoretical biological mechanism that harvests energy to reset the system to a high potential energy state, allowing the transmission of additional signals. Energy may be induced into the active metamaterial through a biological process, such as actin filaments in eukaryotic cells \cite{Pollard:1986cz,Blanchoin:2014jr,Kumar:2017hw}. That said, our analysis does not necessarily require this mechanism to have a biological origin: the reaction mechanism may also represent a mechanical system where energy is added through other electrochemical \cite{Snita:1997fa}, photovoltaic, thermodynamic \cite{Ding:2017ej} or pneumatic \cite{Wehner:2014be} subsystems.

By finding evidence of travelling wave solutions, we are able to analyse limiting behaviour describing the signal transmission speed and wave shape. We provide a detailed analysis to qualitatively and quantitatively understand the effect of our reaction mechanism on signal transmission abilities of the material.  Our main results consist of a set of analytical approximations that quantify the signal transmission speed as a function of the parameters which describe the physical properties of the material. Results in \reffig{Fig-7-Approximations}{} show that the approximation we develop to apply through the whole domain, given by \refeq{eq:CombinedForm}, provides an excellent match to the numerical results, particularly for large $\eta$. In addition, our approximation for the wavespeed near the slow wave, given by \refeq{eq:IntegrabilityEq_Slow_Curve}, is able to provide excellent information about the shape of $c(\kappa)$ as $c \rightarrow 0$, which we find is difficult to obtain numerically. This approximation is also able to provide a region of the parameter space for which signal transmission can occur, given by $\kappa < 4\nu/(5\eta)$ (\refeq{eq:kappa_star}). This understanding of the effect of our mechanism on the signal transmission speed is useful as it allows our active metamaterial to be tuned to produce desirable new behaviours. For example, our results allow quantification of the trade-off between signal transmission speed and the response time, which is essential for controlling the material.  These insights are also essential for building a material containing a biological mechanism that induces energy into the system. Decreasing $\nu$ and $\eta$ in the same proportion increases the transmission speed at the cost of increased sensitivity to noise-induced misfiring, but may be essential if the energy budget is small.

A key aspect of our study is to follow Nadkarni \etal \cite{Nadkarni2016} by representing the bistable potential energy function as a quartic (\refeq{eq:DiscreteODE-V}).  This approach leads us to obtain numerous analytical approximations that characterise the effect of the biological mechanism on the transmission speed which, although qualitatively reliable, may not always be quantitatively appropriate for particular systems \cite{Raney2016}. In fact, the analytical expression for the transmission speed is a result of the similarity between our model and the well-studied bistable equation \cite{MathematicalPhysiologyI}. These choices mean that our system has mechanical and algebraic properties that are similar to other bistable systems, such as the FitzHugh-Nagumo model \cite{FitzHugh1955,Beck:2008bx}. That said, we do not assume that the timescale of the response is significantly slower than the timescale of the excitement, as is often the case in analysis of such models. Indeed, our aim is to develop an intelligent biomechanical material that has tuneable properties. In some sense, it is desirable that the response is as fast as possible to allow for a short period of time between signal reception and retransmission.  Future work may examine the role of heterogeneities in the properties of the material \cite{Hwang:2018ju,Murphy:2019ee}. Such features could allow the material to selectively transmit signals by creating energy barriers that interact with signals of certain properties \cite{Fang:2017hi} .

The travelling wave analysis we conduct assumes a material of infinite length over a large period of time. However, applications of our material will a have finite length and may have properties not suitable for a continuum model. For example, in a material where the spacing between elements is not significantly different to the length of the material, a discrete travelling wave analysis may be more appropriate. The discrete  problem is known to  be substantially more difficult that the continuous problem \cite{MathematicalPhysiologyI}, so the limiting transmission speed our analysis provides may still be useful. Furthermore, the inclusion of our biologically inspired mechanism can be incorporated into passive metamaterials of higher dimensions to enable new behaviours and the travelling wave analysis can be extended to investigate two-dimensional signal propagation.  In the supporting material we produce results which show the transmission of concurrent signals (Figure S1 and S2) and interacting signals initiated from both ends of the material (Figure S3). Further analysis is needed to examine the behaviour of these types of interacting waves \cite{Simpson:2007dz} and the material's ability transmit oscillatory or concurrent signals. 

To conclude, we have presented a novel, biologically inspired, active metamaterial that can reversibility transmit mechanical signals. This work provides an analytical expression that describes the mechanical properties of the material required for signal transmission. We also provide numerous approximations that quantify the effect of the mechanical properties, and the timescale of the biological response, on the transmission speed. This work demonstrates how a new class of biologically inspired metamaterials are able to produce useful new functionalities. The type of analysis we present is invaluable for tuning and controlling the active metamaterial.

\vspace{20pt}

\noindent\textbf{Data Accessibility}\\
\noindent This article has no additional data. Key algorithms used to generate results are available on Github at \texttt{\href{https://github.com/ap-browning/rspa-2019}{github.com/ap-browning/rspa-2019}}.\\

\noindent\textbf{Contributions}\\
\noindent All authors conceived and designed the study; APB performed the analysis and numerical simulations, and drafted the article; all authors provided comments and gave final approval for publication.\\

\medskip
\noindent\textbf{Funding}\\
\noindent This work is supported by the Australian Research Council (DP170100474) and the Royal Society exchanges grant no. IE160805.\\

\medskip
\noindent\textbf{Acknowledgements}\\
\noindent We thank Kevin Burrage and Ian Turner for their helpful discussions. We also thank the three anonymous referees for their comments.


\end{document}


\title{Supplementary Material for\\``Reversible signal transmission in an active mechanical metamaterial''}

\author[1,2]{Alexander P Browning}
\author[3]{Francis G Woodhouse}
\author[1\footnote{Corresponding author. E-mail: matthew.simpson@qut.edu.au}]{Matthew J Simpson}
\affil[1]{Mathematical Sciences, Queensland University of Technology, Brisbane, Australia}
\affil[2]{ARC Centre of Excellence for Mathematical and Statistical Frontiers, QUT, Australia}
\affil[3]{Mathematical Institute, University of Oxford, Oxford, UK}

\maketitle

\tableofcontents
\clearpage

\medskip\section{Numerical solution of the continuous model}
\label{s:NumericalSolutionContinuousModel}

In this section, we explain the numerical scheme used to solve the continuous model in the main paper. The non-dimensional continuous model is given by
	\begin{equation}\label{eq:ContinuousModel}
	\begin{aligned}
		0 &= \pdv[2]{\hat{u}}{\hat{t}} - \pdv[2]{\hat{u}}{\hat{x}} + \pdv{\hat{u}}{\hat{t}} + \nu(\hat{u}^2-1)\left(\frac{\hat{u}}{\eta} -\hat{a}\right),\\
		0 &= \pdv{\hat{a}}{\hat{t}} - \kappa\left(\dfrac{\hat{u}}{\eta} - \hat{a}\right),
	\end{aligned}
	\end{equation}
where $\hat{x} \in (0,L)$, with no-flux boundary conditions on $\hat{x} \in \{0,\hat{L}\}$, and initial conditions given by
	\begin{equation}\label{eq:ContinuousModelIC}
	\begin{aligned}
		\hat{u}(\hat{x},0) &= \left\{\begin{array}{rl}
						1, & 0 \le \hat{x} < Q,\\
						-1, & Q \le \hat{x} \le \hat{L},
					\end{array}\right.\\
		\hat{a}(\hat{x},0) &= \left\{\begin{array}{rl}
				1/\eta, & 0 \le \hat{x} < Q,\\
				-1/\eta, & Q \le \hat{x} \le \hat{L}.
			\end{array}\right.	
	\end{aligned}
	\end{equation}

To solve the system given by \refeq{eq:ContinuousModel}, we employ a finite difference technique. We first introduce
	\begin{equation}
		\hat{w} = \pdv{\hat{u}}{\hat{t}},	
	\end{equation}
so that the system in \refeq{eq:ContinuousModel} can be written as the first order system of partial differential equations (PDEs),
	\begin{equation}
	\begin{aligned}
		\pdv{\hat{u}}{t} &= \hat{w},\\
		\pdv{\hat{w}}{t} &= \pdv[2]{\hat{u}}{\hat{x}} - \hat{w} - \nu(\hat{u}^2 - 1)(\hat{u} - \hat{a}),\\
		\pdv{\hat{a}}{\hat{t}} &= \kappa\left(\dfrac{\hat{u}}{\eta} - \hat{a}\right).
	\end{aligned}
	\end{equation}
Initially, the system is at rest, so that
	\begin{equation}\label{eq:wIC}
		\hat{w}(\hat{x},0) = 0.	
	\end{equation}

Next, we divide the domain, $\hat{x} \in [0,\hat{L}]$ into $S$ equally spaced subintervals, each of width $\delta\hat{x} = \hat{L} / S$. We index the boundary of each subinterval with $n \in \{0,1,...,S\}$, such that $\hat{u}_n(t) \approx \hat{u}(n\delta\hat{x},\hat{t})$. Doing this, we obtain the following system of first order, non-linear, ordinary differential equations,
	\begin{equation}\label{eq:S-PDE-Discretised}
	\begin{aligned}
		\dv{\hat{u}_n}{t} &= \hat{w}_n,\:\forall n,\\
		\dv{\hat{w}_0}{t} &= \frac{\hat{u}_{1}-\hat{u}_0}{\Delta x^2} -  \hat{w}_1 - \nu(\hat{u}_0^2 - 1)(\hat{u}_0-\hat{a}_0),\\
		\dv{\hat{w}_n}{t} &= \frac{\hat{u}_{n+1}-2\hat{u}_n+\hat{u}_{n-1}}{\Delta x^2} -  \hat{w}_n - \nu(\hat{u}_n^2 - 1)(\hat{u}_n-\hat{a}_n),\: 1\le n \le S-1,\\
		\dv{\hat{w}_{S}}{t} &= \frac{\hat{u}_{S}-\hat{u}_{S-1}}{\Delta x^2} - \hat{w}_{S} - \nu(\hat{u}_S^2 - 1)(\hat{u}_S-\hat{a}_S),\\
		\dv{a_i}{t} &= \kappa\left(\dfrac{\hat{u}_n}{\eta} - \hat{a}_n\right)\:\forall n.
	\end{aligned}
	\end{equation}

Next, we let $\textbf{X}(t) = \langle \hat{u}_0,...,\hat{u}_{S},\hat{w}_0,...,\hat{w}_{S},\hat{a}_0,...,\hat{a}_{S}\rangle$ such that \refeq{eq:S-PDE-Discretised} can be written as $\textbf{X}' = \textbf{f}(\textbf{X})$. We can then integrate \refeq{eq:S-PDE-Discretised} using a second order trapezoidal rule,
	\begin{equation}\label{eq:S-PDE-Trap}
		\dfrac{\textbf{X}(t + \Delta t) - \textbf{X}(t)}{\Delta t} \approx \dfrac{1}{2}\bigg(\textbf{f}[\textbf{X}(t + \Delta t)] + \textbf{f}[\textbf{X}(t)] \bigg).
	\end{equation}
We solve \refeq{eq:S-PDE-Trap} using Picard iteration \cite{Eberhard:1996va}, where $\textbf{X}(t)$, which is always known, is taken to be the initial guess at each timestep. The initial conditions are chosen according to \refeqs{eq:ContinuousModelIC}{eq:wIC}.

Unless specified otherwise, results in the main paper are produced using a discretisation where $\delta t = 0.01$ and $S = 10000$.


\medskip\section{Signal initiation}
\label{s:SignialInitiation}

\begin{figure}[!htp]
	\centering
	\includegraphics[scale=0.8]{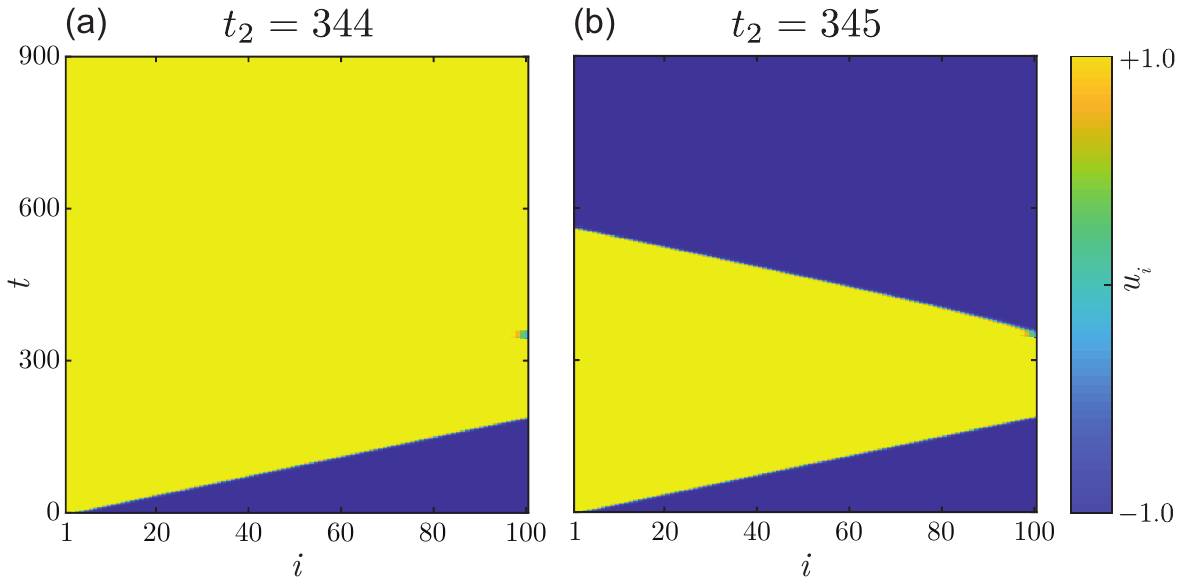}
	\caption{Signal propagation through the material described by the discrete model showing $u_i(t)$, the displacement, where $i$ is the mass index. The first signal was initiated by moving the first element from a displacement of $-\delta$ to $\delta$ at $t = 0$, and was retransmitted by moving the last element from a displacement of $\delta$ to $-\delta$ at $t = t_2$. In (a) the second signal was not able to propagate for this value of $\epsilon$, in (b) the signal is able to propagate. Parameters used are $m = 1$ g, $k = 1$ g m/s$^2$, $\gamma = 1$ g/s, $\Delta = 0.002$ m, $\delta = 1$ m, $\epsilon = 0.01$ /s, $\eta = 2$, $v = 1$ g/(m$^2$s$^2$) and $N = 101$ masses.}
	\label{S1-SecondSignalR}
\end{figure}

\begin{figure}[!htp]
	\centering
	\includegraphics[scale=0.8]{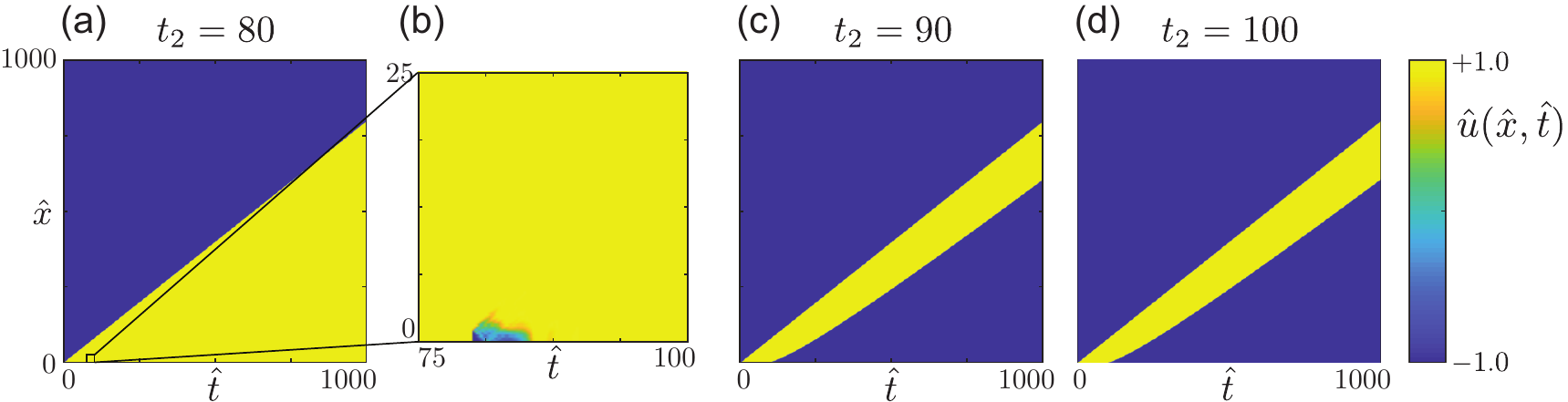}
	\caption{The solution to the continuous model showing $\hat{u}(\hat{x},\hat{t})$. The first signal is initiated by setting $\hat{u}(\hat{x},\hat{t}) = 1$ for $\hat{x} < 1$ at $t = 0$. The second signal was initiated by setting $\hat{u}(\hat{x},\hat{t}) = -1$ for $\hat{x} < 1$ at $t = t_2$. In (a) the second signal is not able to propagate (signal initiation is shown in the inset (b)). In (c) and (d), the second signal is able to propagate.}
	\label{S2-SecondSignalL}
\end{figure}

\begin{figure}[!htp]
	\centering
	\includegraphics[scale=0.8]{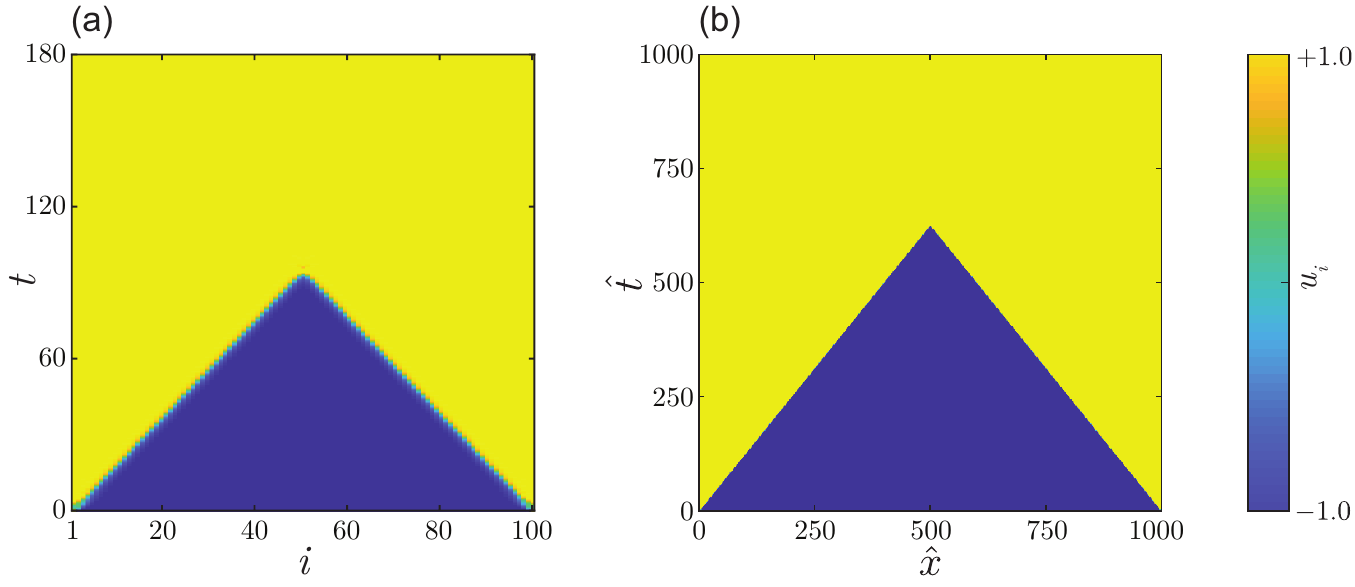}
	\caption{(a) The solution to the discrete model showing showing $u_i(t)$, the displacement, where $i$ is the mass index. A signal is initiated simultaneously at each end of the domain by moving the first and last element to $u_i = 1$ at $t = 0$. (b) The solution to the continuous model showing the displacement field, $\hat{u}(\hat{x},\hat{t})$. A signal is initiated simultaneously at each end of the domain by setting $\hat{u}(\hat{x},\hat{t}) = -1$ for $\hat{x} \le 1$ and $\hat{x} \ge 499$ at $t = 0$. The parameters used in the discrete model are $m = 1$ g, $k = 1$ g m/s$^2$, $\gamma = 1$ g/s, $\Delta = 0.002$ m, $\delta = 1$ m, $\epsilon = 0.01$ /s, $\eta = 2$, $v = 1$ g/(m$^2$s$^2$) and $N = 101$ masses. The parameters used in the continuous model are $\nu = 4$, $\eta = 2$ and $\kappa = 0.01$.}
	\label{S3-Colliding}
\end{figure}

In the main text, we focus on the materials ability to transmit signals that are initiated at the left boundary.  In addition to this, results shown in the main text allow sufficient time between signal reception at the right boundary and retransmission. In \reffig{S1-SecondSignalR}{} we demonstrate how a second signal initiated from the right boundary at $t_2 = 344$ is unable to propagate, whereas a second signal initiated from the right boundary at $t_2 = 345$ is able. In \reffig{S2-SecondSignalL}{} we demonstrate that a second signal initiated from the left boundary at $t_2 = 80$ is unable to propagate, whereas those initiated at $t_2 = 90$ and $t_2 = 100$ are able, albeit with an initially slower transmission speed. Finally, in \reffig{S3-Colliding}{} we show how two signals initiated simultaneously from both the left and right boundaries collide.

\medskip\section{Wavespeed approximation technique}
\label{s:WavespeedApproximation}

In this section, we detail the technique we use to approximate the wavespeed, $c$, from the solution of the continuous model. The technique used to solve the continuous model numerically is described in \refelement{section}{s:NumericalSolutionContinuousModel}. The numerical parameters used to solve the continuous model: for the purpose of estimating the wavespeed in table 1 in the main paper, are shown in \refelement{table}{tab:MainTableParams}; and, for the purpose of estimating the wavespeed in figure 7 in the main paper, are shown in \refelement{table}{tab:FigureParams}.

\begin{figure}[!htp]
	\centering
	\includegraphics[width=\textwidth]{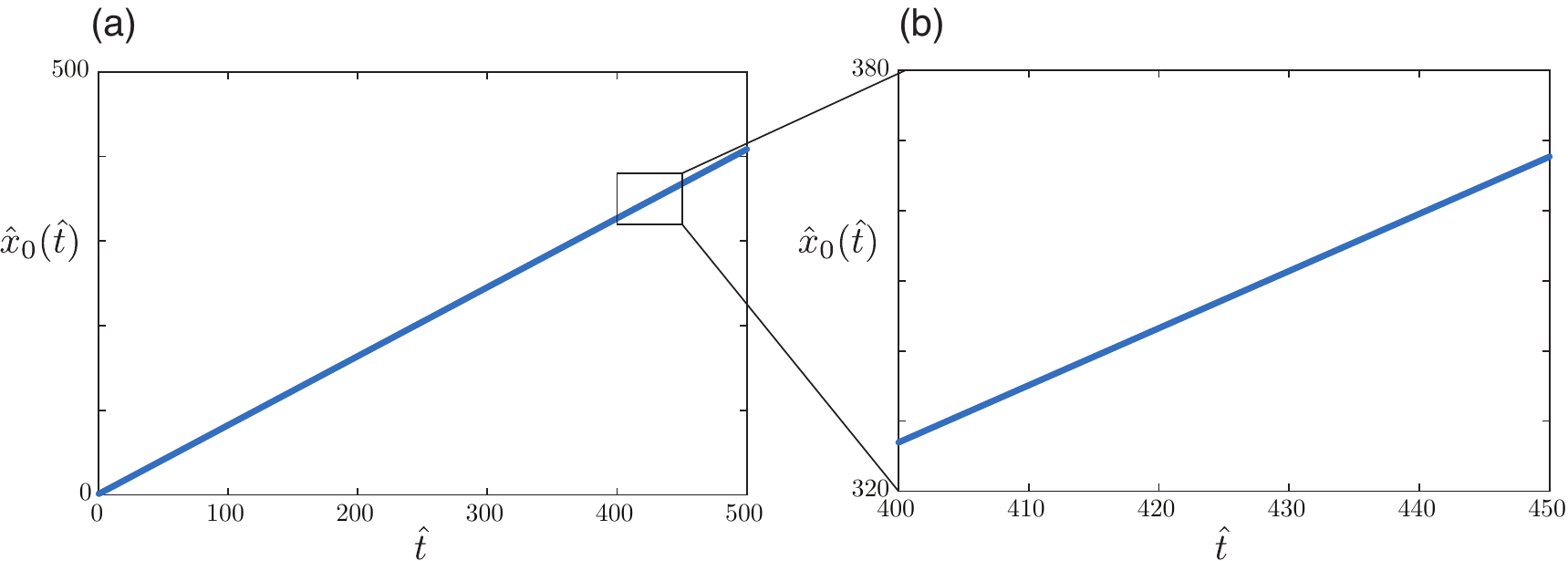}
	\caption{Example wave front time series data, produced using a solution to the continuous model for $\nu = 4$, $\eta = 2$ and $\kappa = 0$. Numerical parameters are detailed in \refelement{table}{tab:MainTableParams}. (a) Shows the front data for $\hat{t} \in [0,\hat{t}_\text{max}]$. (b) Shows the inset of (a) where $\hat{t} \in (\hat{t}_\text{lower},\hat{t}_\text{upper})$, which corresponds to data from which the wavespeed is approximated from.}
	\label{Fig-S4-WavespeedApprox}
\end{figure}

We first convert space-time data to wave-location time series data by defining the location of the front, $x_0(t)$, where
	\begin{equation}
		\hat{x}_0(t) = \hat{x} \: : \: \hat{u}(\hat{x},\hat{t}) = 0.
	\end{equation}
As our data is typically discrete, we find $x_0(t)$ using a linear interpolation. We now have discrete data for $\hat{x}_0(t)$ where $0 \le \hat{t} \le \hat{t}_\text{max}$. An example of this data is shown in \reffig{Fig-S4-WavespeedApprox}{a}.

Next, we determine a section of the wavefront curve that is at ``late time'' and sufficiently far from the right boundary. In this study, we found an appropriate region, to be automatically determined as
	\begin{equation}
		\hat{t} \in (\hat{t}_\text{lower},\hat{t}_\text{upper}) = (0.8\hat{t}_\text{max},0.9\hat{t}_\text{max}).
	\end{equation}
This process gives us (approximately) linear data of the form $\{t^{(i)},x^{(i)}_0\}$, where the slope is the estimate of the wavespeed, $c$. An example of this (approximately) linear data is shown in \reffig{Fig-S4-WavespeedApprox}{b}.

To approximate $c$, we linearly interpolate the data $\{t^{(i)},x^{(i)}_0\}$, and choose $c$ to be the gradient of the regression line.

\begin{table}[!htp]
\centering
\begin{tabular}{C{1cm}C{1cm}|C{1cm}C{1cm}C{1cm}C{1cm}}\hline
$\eta$ & $\nu$ & $\hat{t}_\text{max}$ & $\hat{x}_\text{max}$ & $\delta \hat{t}$  & $S$     \\\hline\hline
1.5  & 1    & 500 & 500  & 0.01 & 10000 \\
1.5  & 2    & 500 & 500  & 0.01 & 10000 \\
1.5  & 4    & 500 & 500  & 0.01 & 10000 \\\hline
2    & 1    & 500 & 500  & 0.01 & 10000 \\
2    & 2    & 500 & 500  & 0.01 & 10000 \\
2    & 4    & 500 & 500  & 0.01 & 10000 \\\hline
3    & 1    & 500 & 500  & 0.01 & 10000 \\
3    & 2    & 500 & 500  & 0.01 & 10000 \\
3    & 4    & 500 & 500  & 0.01 & 10000 \\\hline
4    & 1    & 500 & 2000 & 0.01 & 20000 \\
4    & 2    & 500 & 2000 & 0.01 & 20000 \\
4    & 4    & 500 & 2000 & 0.01 & 20000 \\\hline
\end{tabular}
\caption{Numerical parameters used to solve the continuous model when estimating the wavespeed for $\kappa \ll 1$ in \refelement{table}{tab:cCorrectionEstimates} of this supporting material document.}
\label{tab:MainTableParams}
\end{table}

\begin{table}[!htp]
\centering
\begin{tabular}{C{1cm}C{1cm}|C{1cm}C{1cm}C{1cm}C{1cm}}\hline
$\eta$ & $\nu$ & $\hat{t}_\text{max}$ & $\hat{x}_\text{max}$ & $\delta \hat{t}$  & $S$     \\\hline\hline
1.5  & 1    & 500 & 10000  & 0.01 & 10000 \\
1.5  & 2    & 500 & 1000  & 0.01 & 10000 \\
1.5  & 4    & 500 & 1000  & 0.01 & 10000 \\\hline
2    & 1    & 500 & 10000  & 0.01 & 10000 \\
2    & 2    & 500 & 1000  & 0.01 & 10000 \\
2    & 4    & 500 & 1000  & 0.01 & 10000 \\\hline
3    & 1    & 500 & 10000  & 0.01 & 10000 \\
3    & 2    & 500 & 1000  & 0.01 & 10000 \\
3    & 4    & 500 & 1000  & 0.01 & 10000 \\\hline
4    & 1    & 500 & 10000 & 0.01 & 20000 \\
4    & 2    & 500 & 10000 & 0.01 & 20000 \\
4    & 4    & 500 & 10000 & 0.01 & 20000 \\\hline
\end{tabular}
\caption{Numerical parameters used to solve the continuous model when estimating the wavespeed for $\kappa > 0$ in figure 7 in the main document.}
\label{tab:FigureParams}
\end{table}

\medskip\section{Perturbation solution}
\label{s:PerturbationSolution}

In this section, we present our method for finding a singular perturbation expansion to the travelling wave model, given by the system
	\begin{align}
		0 	&= \dv[2]{f}{z} + \dfrac{c}{1 - c^2}\dv{f}{z} - \dfrac{\nu}{1 - c^2}(f^2 - 1)(f - h),\:\:f(\pm\infty) = \mp 1,\label{eq:TravellingWaveModel}\\
		0 	&= \dv{h}{z} + \dfrac{\kappa}{c}\left(\dfrac{f}{\eta} - h\right),\:\:h(\infty) = -\dfrac{1}{\eta},\label{eq:TravellingWaveModel-H}
	\end{align}

We pose a perturbation expansion for the wavespeed,
	\begin{equation}\label{eq:PertubationExpansion_C}
		c = c_0 + c_1\kappa + \bigok{2},
	\end{equation}
where $c_0$ is known. We take time to note that, for this expansion, we also have 
	\begin{align*}
		\dfrac{1}{c} &= \dfrac{1}{c_0} - \dfrac{c_1}{c_0^2}\kappa + \bigok{2}.\\
		\dfrac{c}{1 - c^2} &= \dfrac{c_0}{1-c_0^2} + \left(\dfrac{c_1}{1-c_0^2} + \dfrac{2c_0^2c_1}{(1-c_0^2)^2}\right)\kappa + \bigok{2},\\
		\dfrac{\nu}{1 - c^2} &= \nu\left(\dfrac{1}{1-c_0^2} + \dfrac{2 c_0c_1}{(1-c_0^2)^2}\kappa + \bigok{2}\right).
	\end{align*}

We now proceed to find perturbation solutions in the inner and outer regions. We denote solutions in the inner region $z \sim \mathcal{O}(\mu_0^{-1})$ using lowercase variables, $f(z)$ and $h(z)$. We denote solutions in the outer region $Z = \kappa z \sim \mathcal{O}(\kappa^{-1})$ using uppercase variables, $F^{(L)}(Z)$, $F^{(R)}(Z)$, $H^{(L)}(Z)$ and $H^{(R)}(Z)$, where a superscript, $(L)$ and $(R)$, denotes solutions in left, $Z < 0$ and right, $Z > 0$ regions of the outer region, respectively.

\vspace{0.3cm}
\noindent\textit{Outer Region}\\
\noindent Substituting $Z = \kappa z$ into \refeqs{eq:TravellingWaveModel}{eq:TravellingWaveModel-H} leads to a system described by
	\begin{equation}\label{eq:S-TW-SlowODE}
		\begin{aligned}
			&\begin{aligned}
				0 	&= \kappa^2\dv[2]{F}{Z} + \dfrac{c\kappa}{1- c^2}\dv{F}{Z} - \dfrac{\nu}{1- c^2}(F^2-1)(F-H),\\
				0 	&= \dv{H}{Z} + \dfrac{1}{c}\left(\dfrac{F}{\eta} - H\right),\\
			\end{aligned}\\
		\end{aligned}
	\end{equation}
The left and right solutions in the outer region will match the corresponding left and right boundary conditions from \refeqs{eq:TravellingWaveModel}{eq:TravellingWaveModel-H}, such that
	\begin{align}
		&F^{(L)}(-\infty) = 1,\: &H^{(L)}(-\infty) = \frac{1}{\eta},\\
		&F^{(R)}(\infty) = -1,\: &H^{(R)}(\infty) = -\frac{1}{\eta}.	
	\end{align}
Provided $\kappa \ll \mu_0$, for $Z \sim \mathcal{O}(1)$ the solution to the system for $\kappa = 0$ is given by 
	\begin{equation*}
		f\left(\dfrac{Z}{\kappa}\right) = \tanh\left(\dfrac{\mu_0}{\kappa}Z\right) \sim -\text{sign}(Z),
	\end{equation*}
since $\mu_0 / \kappa \ll 1$. It is then trivial to see that the expansion for $F(Z)$ contains only the leading order term, so that
	\begin{equation}
		F(Z) = -\text{sign}(Z) \Rightarrow \begin{array}{ll}F^{(L)}(Z) = 1,& Z < 0,\\F^{(R)}(Z) = -1,&Z > 0.\end{array}
	\end{equation}
Therefore, the solution for $H(Z)$ where $Z > 0$, $H^{(R)}(Z)$, is given by
	\begin{equation}
		H^{(R)}(Z) = -\frac{1}{\eta}.
	\end{equation}

To find the solution for $H(Z)$ where $Z < 0$, $H^{(L)}(Z)$, we pose a perturbation solution of the form
	\begin{equation}
		H^{(L)}(Z) = H_0 + H_1\kappa + \bigok{2}.
	\end{equation}
Substituting this perturbation expansion into \refeq{eq:S-TW-SlowODE} gives
	\begin{equation}\label{eq:HLexpansion}
		0 = H_0' + \dfrac{1}{c_0}\left(\dfrac{1}{\eta} - H_0\right) + \left[H_1' - \dfrac{1}{c_0}H_1 + \dfrac{c_1}{c_0^2}\left(\dfrac{1}{\eta} - H_0\right)\right]\kappa + \bigok{2},\\
	\end{equation}
where the boundary conditions are given by 
	\begin{equation}
		\lim_{Z \rightarrow -\infty} = \left\{\begin{array}{ll}
					-\dfrac{1}{\eta}, & i = 0,\\
					0, & i > 1.
				\end{array}\right.	
	\end{equation}

The $\mathcal{O}(1)$ term in \refeq{eq:HLexpansion} gives the separable ordinary differential equation
	\begin{equation}
			0 = H_0' + \dfrac{1}{c_0}\left(\dfrac{1}{\eta} - H_0\right),\:Z < 0,\:\lim_{Z\rightarrow -\infty}H_0(Z) = \frac{1}{\eta},
	\end{equation}
which has the solution
	\begin{align*}
		H_0(Z) &= \alpha_1 \exp\left(\dfrac{Z}{c_0}\right) + \dfrac{1}{\eta},
	\end{align*}
where $\alpha_1$ is to be determined by matching solutions in the inner and outer regions to $\mathcal{O}(1)$.

The $\bigok{}$ term in \refeq{eq:HLexpansion} gives the ordinary differential equation
	\begin{equation}\label{eq:fast_ode_barred}
		0 = H_1' - \dfrac{1}{c_0}H_1 + \dfrac{c_1}{c_0^2}\alpha_1\exp\left(\dfrac{Z}{c_0}\right),\:Z < 0,\:\lim_{Z \rightarrow -\infty} H_1(Z) = 0,
	\end{equation}
which has the solution
	\begin{equation}
		H_1(Z) = \left(\alpha_2 - \dfrac{c_1}{c_0^2}\alpha_1Z\right)\exp\left(\dfrac{Z}{c_0}\right),
	\end{equation}
where $\alpha_2$ is to be determined by matching solutions in the inner and outer regions to $\bigok{}$.

\vspace{0.3cm}
\noindent\textit{Inner Region}\\
In the inner region, we pose a perturbation solution of the form
	\begin{align}
		f(z) &= f_0(z) + f_1(z) \kappa + \bigok{2},\label{eq:InnerExpansion-F}\\
		h(z) &= h_0(z) + h_1(z) \kappa + \bigok{2},\label{eq:InnerExpansion-H}
	\end{align}
where $f_0$ and $h_0$ are the solutions for $\kappa = 0$,
	\begin{align}
		f_0(z) &= -\tanh(\mu_0 z),\\
		h_0(z) &= -\dfrac{1}{\eta}.
	\end{align}

Substituting \refeqs{eq:InnerExpansion-F}{eq:InnerExpansion-H} into \refeqs{eq:TravellingWaveModel}{eq:TravellingWaveModel-H} gives
	\begin{align}
			&\begin{aligned}
			0 &= \dv[2]{f_0}{z} + \dfrac{1}{c_0}\dv{f_0}{z} - \dfrac{\nu}{1 - c_0^2}(f_0^2-1)(f_0-h_0)\alignnl
				+ \bigg\{\dv[2]{f_1}{z} + \dfrac{1}{c_0}\dv{f_1}{z} -\dfrac{\nu}{1 - c_0^2}\bigg(2f_0(f_0-h_0) + (f_0^2-1)\bigg)f_1 \alignnl
				+ \left(\dfrac{c_1}{1-c_0^2} + \dfrac{2c_0^2c_1}{(1-c_0^2)^2}\right) \dv{f_0}{z} - \dfrac{2\nu c_0c_1}{(1-c_0^2)^2} (f_0^2-1)(f_0-h_0) \alignnl
				+ \dfrac{\nu}{1 - c_0^2}(f_0^2-1)h_1\bigg\}\kappa + \bigok{2},
			\end{aligned}\label{eq:fast_ode_expansion_f}\\	
			&\begin{aligned}
			0 &= \dv{h_0}{z}%
				+ \bigg\{\dv{h_1}{z} + \dfrac{1}{c_0}\left(\dfrac{f_0}{\eta} - h_0\right)\bigg\}\kappa + \bigok{2}.
			\end{aligned}\label{eq:fast_ode_expansion_h}
	\end{align}

The $\bigok{1}$ term in \refeq{eq:fast_ode_expansion_h} is directly integrable. Consider 
	\begin{equation}
		\begin{aligned}
			0 = \dv{h_1}{z} + \dfrac{1}{c_0}\left(\dfrac{f_0}{\eta} - h_0\right),\\
		\end{aligned}
	\end{equation}
which has solution 
	\begin{equation}\label{eq:fast_ode_sol_kappa_h}
		h_1(z) = \dfrac{1}{c_0 \mu_0 \eta} \bigg(\log(\cosh(\mu_0 z)) - \mu_0 z+ \alpha_3\bigg),
	\end{equation}
where $\alpha_3$ is to be determined by matching solutions in the inner and outer regions to $\bigok{}$.

To solve for $f_1(z)$, we consider that the $\bigok{1}$ term in \refeq{eq:fast_ode_expansion_f} can be written as
	\begin{equation}\label{eq:S-BVP_f1}
		0 = \dv[2]{f_1}{z} + p(z)\dv{f_1}{z} + q(z)f_1 + c_1r(z) + s(z),\:\:\lim_{z\rightarrow \pm\infty} f_1(z) = 0,
	\end{equation}
where $c_1$ is a yet undetermined correction term to the wave-speed, and
	\begin{equation}
		\begin{aligned}
			p(z) &= \dfrac{1}{c_0},\\
			q(z) &= -\dfrac{\nu}{1-c_0^2} \bigg(3\tanh^2(\mu_0 z) - \dfrac{2}{\eta}\tanh(\mu_0 z) - 1\bigg),\\
			r(z) &= -\dfrac{\sech^2(\mu_0 z)}{1 - c_0^2}\left[\mu_0\left(1 + \dfrac{2c_0^2}{1 - c_0^2}\right) + \dfrac{2c_0\nu}{1 - c_0^2}\left(\tanh(\mu_0 z) - \dfrac{1}{\eta}\right)\right],\\
			s(z) &= -\dfrac{\nu\sech^2(\mu_0 z)}{(1 - c_0^2)c_0\mu_0Okap\eta}\left(\log(\cosh(\mu_0 z)) - \mu_0 z + \log(2)\right).
		\end{aligned}
	\end{equation}
The boundary conditions are chosen so that $f_1(z)$ is able to match the correction term in the outer solution, $F_1(Z) \equiv 0$. We describe the numerical technique we use to solve this boundary value problem in \refelement{section}{s:BVP} of this supporting material document.

\vspace{0.3cm}
\noindent\textit{Matching}\\
Here, we apply van Dyke's matching rule \cite{Hinch1991} to determine the integration constants that have appeared in each solution.

First, we match the fast solution to the slow right solution to $\bigok{}$,
	\begin{align*}
		\lim_{\kappa \rightarrow 0} H^{(R)}(\kappa z) &= \lim_{\kappa \rightarrow \infty}\left[ h_0\left(\dfrac{Z}{\kappa}\right) + h_1\left(\dfrac{Z}{\kappa}\right)\kappa\right],\\
		-\frac{1}{\eta} &= -\frac{1}{\eta} + \lim_{\kappa \rightarrow \infty} \dfrac{1}{c_0 \mu \eta} \bigg(\log(\cosh(\dfrac{\mu}{\kappa}Z)) - \dfrac{\mu}{\kappa}Z+ \alpha_3 \bigg) ,\\
		\Rightarrow \alpha_3 &= \log(2).
	\end{align*}

Next, we match the fast solution to the slow left solution to $\mathcal{O}(1)$,
	\begin{align*}
		\lim_{\kappa \rightarrow 0} H_0(\kappa z) &= \lim_{\kappa \rightarrow \infty} h_0\left(\dfrac{Z}{\kappa}\right),\\
		\lim_{\kappa \rightarrow 0} \left(\alpha_1 \exp\left(\dfrac{\bar{c}_0\kappa\zeta}{k}\right) + \dfrac{1}{\eta}\right) &= \lim_{\kappa \rightarrow \infty} -\frac{1}{\eta},\\
		\Rightarrow \alpha_1 = -\frac{2}{\eta},
	\end{align*}
and to $\bigok{1}$,
	\begin{align*}
		\lim_{\kappa \rightarrow 0} H_1(\kappa z)\kappa &= \lim_{\kappa \rightarrow \infty}h_1\left(\dfrac{Z}{\kappa}\right)\kappa,\\
		\lim_{\kappa \rightarrow 0} \left(\alpha_2 - \dfrac{2c_1}{\eta c_0^2}\kappa z\right)\exp\left(\dfrac{z}{\kappa c_0}\right)\kappa &= \lim_{\kappa \rightarrow \infty} \dfrac{1}{c_0 \mu \eta} \bigg(\log(\cosh(\dfrac{\mu}{\kappa}Z)) - \dfrac{\mu}{\kappa}Z + \log(2)\bigg)\kappa,\\
		\Rightarrow \alpha_2 = 0.
	\end{align*}

In summary, we have that 
	\begin{align*}
		F(Z) &= -\text{sign}(Z),\\
		f(z) &= -\tanh(\mu_0 z) + f_1(z)\kappa + \bigok{2},\\
		H^{(L)}(Z) &= \dfrac{1}{\eta} - \dfrac{2}{\eta} \exp\left(\dfrac{Z}{c_0}\right) - \dfrac{2c_1}{\eta c_0^2}Z\exp\left(\dfrac{Z}{c_0}\right)\kappa + \bigok{2},\\
		H^{(R)}(Z) &= -\dfrac{1}{\eta},\\
		h(z) &= -\dfrac{1}{\eta} + \dfrac{1}{c_0 \mu_0 \eta} \bigg(\log(\cosh(\mu_0 z)) - \mu_0 z + \log(2)\bigg)\kappa + \bigok{2}.
	\end{align*}
%

\medskip\section{Numerical solution of the boundary value problem}
\label{s:BVP}

In \refelement{section}{s:PerturbationSolution} of this supporting material document, we find that $f_1(z)$ and $c_1$ are described by the boundary value problem
	\begin{equation}\label{eq:S-BVP_f1}
		0 = \dv[2]{f_1}{z} + p(z)\dv{f_1}{z} + q(z)f_1 + c_1r(z) + s(z),\:\:\lim_{z\rightarrow \pm\infty} f_1(z) = 0,
	\end{equation}
where $c_1$ is a yet undetermined correction term to the wavespeed, and
	\begin{equation}
		\begin{aligned}
			p(z) &= \dfrac{1}{c_0},\\
			q(z) &= -\dfrac{\nu}{1-c_0^2} \bigg(3\tanh^2(\mu_0 z) - \dfrac{2}{\eta}\tanh(\mu_0 z) - 1\bigg),\\
			r(z) &= -\dfrac{\sech^2(\mu_0 z)}{1 - c_0^2}\left[\mu_0\left(1 + \dfrac{2c_0^2}{1 - c_0^2}\right) + \dfrac{2c_0\nu}{1 - c_0^2}\left(\tanh(\mu_0 z) - \dfrac{1}{\eta}\right)\right],\\
			s(z) &= -\dfrac{\nu\sech^2(\mu_0 z)}{(1 - c_0^2)c_0\mu_0\eta}\left(\log(\cosh(\mu_0 z)) - \mu_0 z + \log(2)\right).
		\end{aligned}
	\end{equation}

We approximate the solution to the infinite-domain boundary value problem given by \refeq{eq:S-BVP_f1} as the solution to the finite domain problem,
	\begin{equation}
		0 = \dv[2]{Y}{z} + p(z)\dv{Y}{z} + q(z)Y + c_1r(z) + s(z),\: Y(\pm L) = 0,
	\end{equation}
where $f_1(z) = \lim_{L\rightarrow \infty} Y(z)$. We find that $L = 10$ is sufficient to obtain accurate estimates of $f_1(z)$ and $c_1$.

Next, we divide the domain $z \in [-L,L]$ into $S$ equally space subintervals, each of with $\delta z$. We index the boundary of each subinterval with $n \in \{0,1,...,S\}$, such that $Y_n(z) \approx Y(n\delta z)$. Doing this we obtain the following system of linear equations in terms of the unknowns $Y_1,...Y_{S-1},c_1$,
	\begin{equation}
		\begin{aligned}
			-s_n &= \left(\dfrac{1}{\delta z^2} - \dfrac{p_n}{2\delta z}\right)Y_{n-1} + \left(-\dfrac{2}{\delta z^2} + q_n\right)Y_n + \left(\dfrac{1}{\delta z^2} + \dfrac{p_n}{2\delta z}\right)Y_{n+1} + r_nc_1,\\
			&\hspace{1cm} 1\le n \le S-1,\\
		\end{aligned}
	\end{equation}
where $Y_0 = Y_{S} = 0$ enforces the boundary conditions, $s_n = s(n\delta z)$, and similar for $p(z)$, $q(z)$ and $r(z)$. 

Denoting the unknowns $\textbf{X} = [Y_1,...,Y_{S-1},c_1]^T \in \mathbb{R}^{S}$, we may write the system as the under-determined matrix system
	\begin{equation}\label{eq:Axb}
		\textbf{A}\textbf{X} = \textbf{B},	
	\end{equation}
where $\textbf{B} = [s_2,...,s_S]^T \in \mathbb{R}^{S}$ and $\textbf{A} \in \mathbb{R}^{S\times(S-1)}$. We obtain a solution $\textbf{X}$ from the underdetermined system using the Moore-Penrose pseudoinverse \cite{Campbell:1991vs,MATLABpinv} with the \texttt{pinv} command in MATLAB \cite{MATLABpinv}, which gives the solution with the smallest norm of all possible solutions \cite{Campbell:1991vs,MATLABpinv}. We find that the component of the normalised null-space relating to the free parameter, $c_1$, is $\mathcal{O}(10^{-5})$, so solutions to the system of equations that have $Y(Z)$ small have similar values of the free parameter, $c_1$. Because of this, we find that the solution obtained from the pseudoinverse matches numerical results, obtained from the solution to the continuous model, to two significant figures, providing confidence in the approximate solution to the travelling wave obtained from the continuous model.

\newpage
\medskip\section{Comparison of wavespeed calculations}
\label{s:BVP}

%
\begin{table}[!h]
\centering
\begin{tabular}{C{0.5cm} C{0.5cm} C{1.3cm} C{1.3cm} | C{1.3cm} C{1.3cm} C{1.3cm} }
\hline
\multirow{2}{*}{$\bm\eta$} & \multirow{2}{*}{$\bm\nu$} & \multicolumn{2}{c|}{$\bm{c_0}$} & \multicolumn{3}{c}{$\bm{c_1}$} \\ \cline{3-7} 
 &  & \textbf{PDE} & \textbf{Eq (3.5)} & \textbf{PDE} & \textbf{BVP} & \textbf{Eq (3.26)}\\ \hline
1.5 & 1 & 0.69 & 0.69 & $-0.26$ & $-0.26$ & $-0.30$	\\
1.5 & 2 & 0.80 & 0.80 & $-0.10$ & $-0.10$ & $-0.12$	\\
1.5 & 4 & 0.88 & 0.88 & $-0.03$ & $-0.03$ & $-0.04$	\\ \hline
2 & 1   & 0.58 & 0.58 & $-0.43$ & $-0.43$ & $-0.48$	\\
2 & 2   & 0.71 & 0.71 & $-0.20$ & $-0.20$ & $-0.22$	\\
2 & 4   & 0.82 & 0.82 & $-0.07$ & $-0.08$ & $-0.09$	\\ \hline
3 & 1   & 0.43 & 0.43 & $-0.67$ & $-0.67$ & $-0.73$	\\
3 & 2   & 0.55 & 0.55 & $-0.37$ & $-0.37$ & $-0.40$	\\
3 & 4   & 0.69 & 0.69 & $-0.17$ & $-0.17$ & $-0.19$	\\ \hline
4 & 1   & 0.33 & 0.33 & $-0.81$ & $-0.81$ & $-0.86$	\\
4 & 2   & 0.45 & 0.45 & $-0.49$ & $-0.49$ & $-0.52$	\\
4 & 4   & 0.58 & 0.58 & $-0.26$ & $-0.27$ & $-0.28$	\\ \hline
\end{tabular}
\caption{Comparison of calculations of the wavespeed, $c_0$, and the first correction term in a perturbation solution in $\kappa$, about $\kappa = 0$, $c_1$, shown to two significant figures.  Approximations to $c_1$ using the PDE are performed by fitting a fifth-order polynomial through estimates of $c$ obtained for $\kappa \in \{0,0.01,...,0.05\}$. The numerical scheme, including numerical parameters chosen, are outlined in this supporting material document. Equations refer to the main document.}
\label{tab:cCorrectionEstimates}
\end{table}
